\documentclass[12pt,sort&compress]{elsarticle}
\usepackage{amsmath}
\usepackage{amsfonts}
\usepackage{amssymb}
\usepackage{amsthm}
\usepackage{bm}
\usepackage{indentfirst}
\usepackage{graphicx}
\usepackage{subfigure}
\usepackage{array}
\usepackage{fullpage}
\usepackage{color}

\DeclareMathAlphabet{\mathsfsl}{OT1}{cmss}{m}{sl}

\newcommand{\PreserveBackslash}[1]{\let\temp=\\#1\let\\=\temp}
\newcolumntype{C}[1]{>{\PreserveBackslash\centering}p{#1}}
\newcolumntype{R}[1]{>{\PreserveBackslash\raggedleft}p{#1}}
\newcolumntype{L}[1]{>{\PreserveBackslash\raggedright}p{#1}}

\renewcommand{\vartheta}{\Theta}

\definecolor{mygreen}{rgb}{0.1,0.75,0.2}

\numberwithin{equation}{section}
\newtheorem{thm}{Theorem}[section]

\theoremstyle{definition}

\usepackage{hyperref}
\usepackage{fancyvrb}
\usepackage{comment}
\usepackage{fancyhdr}
\biboptions{sort&compress}

\usepackage{algorithm}
\usepackage[noend]{algpseudocode}
\newcounter{algsubstate}
\renewcommand{\thealgsubstate}{\alph{algsubstate}}

\usepackage{mathtools}
\mathtoolsset{showonlyrefs=true}

\usepackage{dsfont}
\usepackage{bm}
\usepackage{mathrsfs}
\usepackage{amsmath,mathtools}
\usepackage{amsfonts}
\usepackage{amsthm}
\usepackage{xcolor}
\usepackage{setspace}
\usepackage{float}
\usepackage{url}
\usepackage{multicol}
\usepackage{subfigure}
\usepackage{bm}
\usepackage[top=1in,bottom=1in,left=1in,right=1in]{geometry}

\definecolor{officegreen}{rgb}{0.0, 0.5, 0.0}

\newcommand{\mcI}{\mathcal{I}}

\newcommand{\mcX}{\mathcal{X}}

\newcommand{\mcL}{\mathcal{L}}

\newcommand{\mbR}{\mathbb{R}}

\newcommand{\mbRn}{{\mathbb{R}^n}}

\def \bb{\mathbf{b}}
\def \db{\mathbf{d}}

\def \Fb{\mathbf{F}}
\def \Bb{\mathbf{B}}
\def \fb{\mathbf{f}}
\def \ub{\mathbf{u}}
\def \Ub{\mathbf{U}}
\def \Vb{\mathbf{V}}
\def \Pb{\mathbf{P}}

\def \vb{\mathbf{v}}
\def \xb{\mathbf{x}}
\def \Xb{\mathbf{X}}
\def \yb{\mathbf{y}}

\def \Db{\mathbf{D}}

\def \zerob{\mathbf{0}}

\def \veps{\varepsilon}

\newcommand{\real}{\mathbb{R}}

\usepackage{multirow}

\begin{document}

\theoremstyle{definition}
\newtheorem{remark}{Remark}

\newcommand{\vertiii}[1]{{\left\vert\left\vert\left\vert #1
    \right\vert\right\vert\right\vert}}
\newcommand{\vertii}[1]{{\left\vert\left\vert #1
    \right\vert\right\vert}}
\newcommand{\verti}[1]{{\left\vert #1
    \right\vert}}    

\begin{frontmatter}

\title{A data-driven peridynamic continuum model\\ for upscaling molecular dynamics}

\address[yy]{Department of Mathematics, Lehigh University, Bethlehem, PA}
\address[ss]{Center for Computing Research, Sandia National Laboratories, Albuquerque, NM}
\address[md]{Computational Science and Analysis, Sandia National Laboratories, Livermore, CA}

\author[yy,md]{Huaiqian You}\ead{huy316@lehigh.edu}
\author[yy]{Yue~Yu\corref{cor1}}\ead{yuy214@lehigh.edu}
\author[ss]{Stewart Silling}\ead{sasilli@sandia.gov}
\author[md]{Marta D'Elia}\ead{mdelia@sandia.gov}

\begin{abstract}
Nonlocal models, including peridynamics, often use integral operators that embed lengthscales in their definition. However, the integrands in these operators are difficult to define from the data that are typically available for a given physical system, such as laboratory mechanical property tests. In contrast, molecular dynamics (MD) does not require these integrands, but it suffers from computational limitations in the length and time scales it can address. To combine the strengths of both methods and to obtain a coarse-grained, homogenized continuum model that efficiently and accurately captures materials' behavior, we propose a learning framework to extract, from MD data, an optimal Linear Peridynamic Solid (LPS) model as a surrogate for MD displacements. To maximize the accuracy of the learnt model we allow the peridynamic influence function to be partially negative, while preserving the well-posedness of the resulting model. To achieve this, we provide sufficient well-posedness conditions for discretized LPS models with sign-changing influence functions and develop a constrained optimization algorithm that minimizes the equation residual while enforcing such solvability conditions. This framework guarantees that the resulting model is mathematically well-posed, physically consistent, and that it generalizes well to settings that are different from the ones used during training. We illustrate the efficacy of the proposed approach with several numerical tests for single layer graphene. Our two-dimensional tests show the robustness of the proposed algorithm on validation data sets that include thermal noise, different domain shapes and external loadings, and discretizations substantially different from the ones used for training.
\end{abstract}


\end{frontmatter}

\tableofcontents

\section{Introduction}
Complex systems where small-scale dynamics and interactions affect the global behavior are ubiquitous in scientific and engineering applications. In disciplines ranging from climate forecasts to material design, heterogeneities in materials and media at the micro or molecular scales need to be accurately captured to guarantee reliable and trustworthy predictions. 
However, higher degrees of complexity and heterogeneity require numerical simulations of classical mathematical models at small scales that cannot be afforded, despite recent advances in computational power. This fact creates the need for new mathematical models that act at larger scales and that, combined with new advanced architectures, allow for fast predictions \cite{zohdi2017homogenization,bensoussan2011asymptotic,weinan2003multiscale,efendiev2013generalized,junghans2008transport,kubo1966fluctuation,santosa1991dispersive,dobson2010sharp,ortiz1987method,moes1999simplified,hughes2004energy}. The process of upscaling models or data hides several pitfalls that may compromise the reliability of the resulting surrogates. 
 
In the presence of heterogeneities, it is often the case that to adequately reproduce the large-scale behavior of a system, a homogenized model must follow different governing laws, as well as different constitutive properties, than the ones that apply at the small scale. Homogenization theory addresses the approximate treatment of a partial differential equation (PDE) that contains small-scale oscillations in its coefficients \cite{milton02}. It seeks to replace these coefficients with constant or slowly-varying coefficients such that the resulting solutions closely approximate solutions to the original problem in an averaged sense. The resulting parameters are called {\emph{effective properties}}. In many theoretical treatments, the effective properties are valid only in the limiting case of a very small length scale in the oscillatory behavior of the original parameters. The classical notion of effective properties therefore ``washes out'' the length scale in the original problem, causing {\it important information to be lost}. 

Nonlocality in the spatial dependence of a continuum model has long been recognized as a consequence of homogenization \cite{eringen1972nonlocal,bobaru2016handbook}. For example, in continuum mechanics, nonlocality arises from taking the ensemble average of the displacement field in a family of random linear elastic heterogeneous materials 
\cite{beran70,cher06,karal64,rahali15,smy00,willis85}.
To some extent, this nonlocality can be incorporated in homogenized {\it weakly nonlocal} PDEs that embed length scales in their coefficients.
However, weakly nonlocal PDEs are generally insufficient to fully reproduce coarse-grained data because of the limited spectrum of processes that they can describe \cite{du20}.  
In general, increasing the accuracy of weakly nonlocal PDEs to match small-scale data involves using 
higher and higher order partial derivatives, resulting in practical challenges in numerical implementations.

As pointed out in \cite{du20}, {\it nonlocal operators} \cite{du2011mathematical,madenci2019peridynamic} are among the best candidates as model descriptions that can circumvent these limitations. Theoretical and numerical techniques for nonlocal models are not as advanced as for classical PDEs. This is one of the main reasons why integral operators historically have not received a broader adoption in the context of numerical homogenization. However, current advances in nonlocal theory, computer power, and solution algorithms are making nonlocal equations viable as practical modeling tools. While integral operators have proved to be successful in several contexts such as mechanics \cite{emmrich2007well,silling2007peridynamic} and turbulence \cite{dileoni2021,pang2020npinns}, they have not been systematically explored for coarse graining, or upscaling, of molecular dynamics (MD) models, for which we propose a new rigorous modeling paradigm.

We stress the fact that with nonlocal operators, constitutive laws take the form of kernels (integrand functions), whose functional form cannot be established a priori. Although the integral constitutive laws must be consistent with the classical effective properties, they contain information about the small-scale response of the system and must be chosen to reproduce this response with the greatest fidelity.
In a few cases, certain forms of nonlocal kernels have been adopted in the engineering community because experimental evidence confirms the efficacy of the model, or because a closed form of integrand that matches desired physical properties can be analytically determined. 
An example of the former situation is fracture mechanics, where peridynamic models have been demonstrated to be accurate \cite{emmrich2007well,silling2007peridynamic,diehl2019review}. Examples of the latter case include those diffusion processes in which the mean square displacement does not exhibit the linear, classical behavior, but instead exhibits an anomalous fractional behavior \cite{Benson2000}. However, at present, only a few preliminary works address the problem of finding an optimal form for the kernel function \cite{xu2020deriving,xu2021learning,You2020Regression,You2021}.

In view of the growing importance of MD as a tool for designing materials with reduced reliance on laboratory testing, we propose to use nonlocal operators as upscaled continuous models for MD displacements. We seek nonlocal models that capture important aspects of the small-scale behavior better than classical homogenization theory. 
Building on our previous works \cite{You2020Regression,You2021} we address the question of how to obtain large-scale nonlocal descriptions that capture MD behavior that would remain hidden in classical approaches to homogenization. To accomplish this, we use machine learning to identify optimal nonlocal kernel functions. The machine learning method is required to perform well with small datasets that may include thermal noise.

We summarize below our main contributions.
\begin{itemize}
\item We identify the {\it best upscaled nonlocal model}, without prior knowledge of the material properties, that accurately describes the material's global behavior based on a small set of possibly noisy data.
\item The optimal nonlocal model is {\it guaranteed to be well-posed} and {\it generalizes well} to settings that are substantially different from the ones used for training. The optimal model is equally accurate for different sources and geometries, so that it enables generalization. 
\end{itemize}

While our ultimate goal is to learn a general integrand for the nonlocal operator, in this work we consider a specific nonlocal model, the Linear Peridynamic Solid (LPS) model \cite{silling_2007} as a first step towards a more general learning tool. We focus our experiments on single layered graphene for which we identify optimal two-dimensional nonlocal models.

\paragraph{Outline of the paper}
Section \ref{sec:smoothing} shows how to obtain, via smoothing functions, a nonlocal model for MD displacements. In Section \ref{sec:lps} summarizes the peridynamic theory, the LPS model, and the discretization technique used in this work.  Section \ref{sec:regression} presents our learning approach include the well-posedness of the learned model by construction. It also provides the algorithmic workflow and implementation details. Section \ref{sec:consistency} illustrates the consistency of the proposed method on manufactured solutions. Section \ref{sec:MD} demonstrates the effectiveness of the learning technique for MD displacements. We illustrate several properties including generalization with respect to loadings, domain size and shape. The effect of thermal noise and the sensitivity of the algorithm to noise intensity are considered. Section \ref{sec:conclusion} summarizes our contributions and provides future research ideas. 

\section{Coarse-graining of molecular dynamics displacements}\label{sec:smoothing}

In this section it is shown how to define an integral, continuous model for a system of particles. 
More details can be found in \cite{Silling2021Chapter}.
Similar results obtained with statistical mechanics can be found in \cite{lehoucq11statmech}. 

Consider an assembly of mutually interacting particles in a crystal with particle mass $M_\veps$, $\veps=1,2,\dots,N$.
Let the reference positions of these particles be $\Xb_\veps$
and their displacement vectors $\Ub_\veps(t)$. 

Suppose that any particle $\gamma$ exerts a force $\Fb_{\veps\gamma}(t)$ on particle $\veps$, 
and set $\Fb_{\veps\veps}=\zerob$. 
These forces are assumed to be antisymmetric: $\Fb_{\gamma\veps}(t)=-\Fb_{\veps\gamma}(t)$,
for all $\veps$, $\gamma$, and $t$. 
It is also assumed that there is a cutoff distance $d$ for the atomic interactions
such that $\Fb_{\veps\gamma}=\zerob$ if $|\Xb_\gamma-\Xb_\veps|>d$.
Each particle $\veps$ is subjected to a prescribed external force $\Bb_\veps(t)$. 

For any continuum material point $\xb\in\mbRn$, define a {\it smoothing} function $\omega(\xb,\cdot)$ such that the following
normalization holds:
\begin{equation}
  \int_\mbRn \omega(\xb,\Xb_\veps)\,d\xb=1
\label{eqn-omnorm}
\end{equation}
for any $\veps$.
For convenience, assume that at any $\xb$, $\omega(\xb,\cdot)$ has compact support over the ball $B_R(\xb)$, for $R>0$. 
Define the smoothed mass density and body force density fields by
\begin{equation}
\rho(\xb)=\sum_{\veps=1}^N \omega(\xb,\Xb_\veps) M_\veps, \qquad \bb(\xb,t)=\sum_{\veps=1}^N \omega(\xb,\Xb_\veps)\Bb_\veps(t),
\label{eqn-densdef}
\end{equation}
and the smoothed displacement field by
\begin{equation}
\ub(\xb,t)=\frac{1}{\rho(\xb)}\sum_{\veps=1}^N \omega(\xb,\Xb_\veps) M_\veps \Ub_\veps(t).
\label{eqn-udef}
\end{equation}
The evolution equation for the smoothed displacements will now be derived.
Newton's second law for the particles has the following form: for any $\veps$
\begin{equation}
M_\veps \ddot\Ub_\veps(t)=\sum_{\gamma=1}^N \Fb_{\veps\gamma}(t)+\Bb_\veps(t).
\label{eqn-newton}
\end{equation}
Differentiating \eqref{eqn-udef} twice with respect to time yields
\begin{equation}
\rho(\xb)\ddot\ub(\xb,t)=\sum_{\veps=1}^N \omega(\xb,\Xb_\veps) M_\veps \ddot\Ub_\veps(t).
\label{eqn-accelx}
\end{equation}
From \eqref{eqn-densdef}, \eqref{eqn-newton}, and \eqref{eqn-accelx},
\begin{equation}\label{eqn-accel2}
\begin{aligned}
\rho(\xb)\ddot\ub(\xb,t)&=&\sum_{\veps=1}^N \omega(\xb,\Xb_\veps) \left[ \sum_{\gamma=1}^N \Fb_{\veps\gamma}(t)+\Bb_\veps(t)\right] \\
&=&\sum_{\veps=1}^N\sum_{\gamma=1}^N \omega(\xb,\Xb_\veps) \Fb_{\veps\gamma}(t)+\bb(\xb,t).
\end{aligned}
\end{equation}
From \eqref{eqn-omnorm} and \eqref{eqn-accel2}, for any $\xb$,
\begin{equation}
\rho(\xb)\ddot\ub(\xb,t)=\sum_{\veps=1}^N\sum_{\gamma=1}^N \omega(\xb,\Xb_\veps) \Fb_{\veps\gamma}(t)
\left[ \int\omega(\yb,\Xb_\gamma)\;d\yb\right]+\bb(\xb,t),
\label{eqn-accel3}
\end{equation}
or, equivalently,
\begin{equation}
\rho(\xb)\ddot\ub(\xb,t)= \int\fb(\yb,\xb,t)\;d\yb+\bb(\xb,t)
\label{eqn-eom}
\end{equation}
where
\begin{equation}
\fb(\yb,\xb,t)=\sum_{\veps=1}^N\sum_{\gamma=1}^N \omega(\xb,\Xb_\veps) \omega(\yb,\Xb_\gamma) \Fb_{\veps\gamma}(t)
\label{eqn-fdef}
\end{equation}
and $\bb$ is given by \eqref{eqn-densdef}.
The properties of $\Fb$ guarantee that the integrand $\fb$ is antisymmetric. 
Since, by assumption, the smoothing functions have support radius $R$ and the interatomic forces have cutoff distance $d$, it follows that
\begin{equation}
    |\yb-\xb|>\delta\quad\implies\quad \fb(\yb,\xb,t)=\mathbf{0}
\label{eqn-mdcutoff}
\end{equation}
for all $t$,
where the {\emph{horizon}} $\delta$ is given by
\begin{equation}
    \delta=2R+d.
\label{eqn-deltadef}
\end{equation}
In summary, defining the displacements and other fields in the continuum description using the smoothing function $\omega$, 
leads directly the nonlocal (or integral) equation of motion \eqref{eqn-eom}.
However, the derivation does not provide a material model, that is, the dependence of $\fb$ on the deformation in terms of the continuum
displacement field $\ub$ is not yet determined.
The goal of the present work is
to identify an optimal form of the integrand function $\fb$ in \eqref{eqn-eom} such that the corresponding nonlocal model 
faithfully represents given MD displacements under a given set of loading conditions on the MD grid.

At finite temperature, thermal oscillations in displacement are present in any MD simulation.
The details of these oscillations are of no interest for purposes of continuum modeling.
However, their net effect on the bulk material properties must be included.
To smooth out the thermal motions while retaining their net effect, a time-smoothing method is applied.
In this method, the following expression is applied to obtain the time-smoothed displacement $\Ub_\veps(t^n)$ of atom $\veps$:
\begin{equation}
  \Ub_\veps(0)=0, \qquad
  \Ub_\veps(t^n) = (1-\iota)\Ub_\veps(t^{n-1})+\iota \widetilde\Ub_\veps(t^n), \quad n>0
    \label{eqn-timesmooth}
\end{equation}
where $\widetilde\Ub_\veps$ is the unsmoothed displacement of atom $\veps$ and $\iota$ is a positive constant, typically on the order of 0.01. 
Then, to obtain displacements that are smoothed in both space and time, the displacement given by \eqref{eqn-timesmooth} is used in \eqref{eqn-udef}:
\begin{equation}\label{eqn:smoothu}
\ub(\xb)=\frac{1}{\rho(\xb)}\sum_{\veps=1}^N \omega(\xb,\Xb_\veps) M_\veps \Ub_\veps(t^F),
\end{equation}
where $t^F$ is the final time of the MD simulation. 
The displacements $\ub(\xb)$ contain noise in the form of spatial fluctuations due to the impossibility of completely smoothing
out all of the thermal oscillations in an MD simulation within a finite simulation time $t^F$, regardless of the value of $\iota$.
The machine learning algorithm described below attempts to extract continuum material properties from the training data even
in the presence of this noise.
In Section \ref{sec:MD}, we will present results on the effectiveness of this ML algorithm in treating this type of noisy training data.


\section{Peridynamics}\label{sec:lps}

\subsection{Peridynamics Background}

\begin{figure}  
\centering
\includegraphics[width=0.6\textwidth]{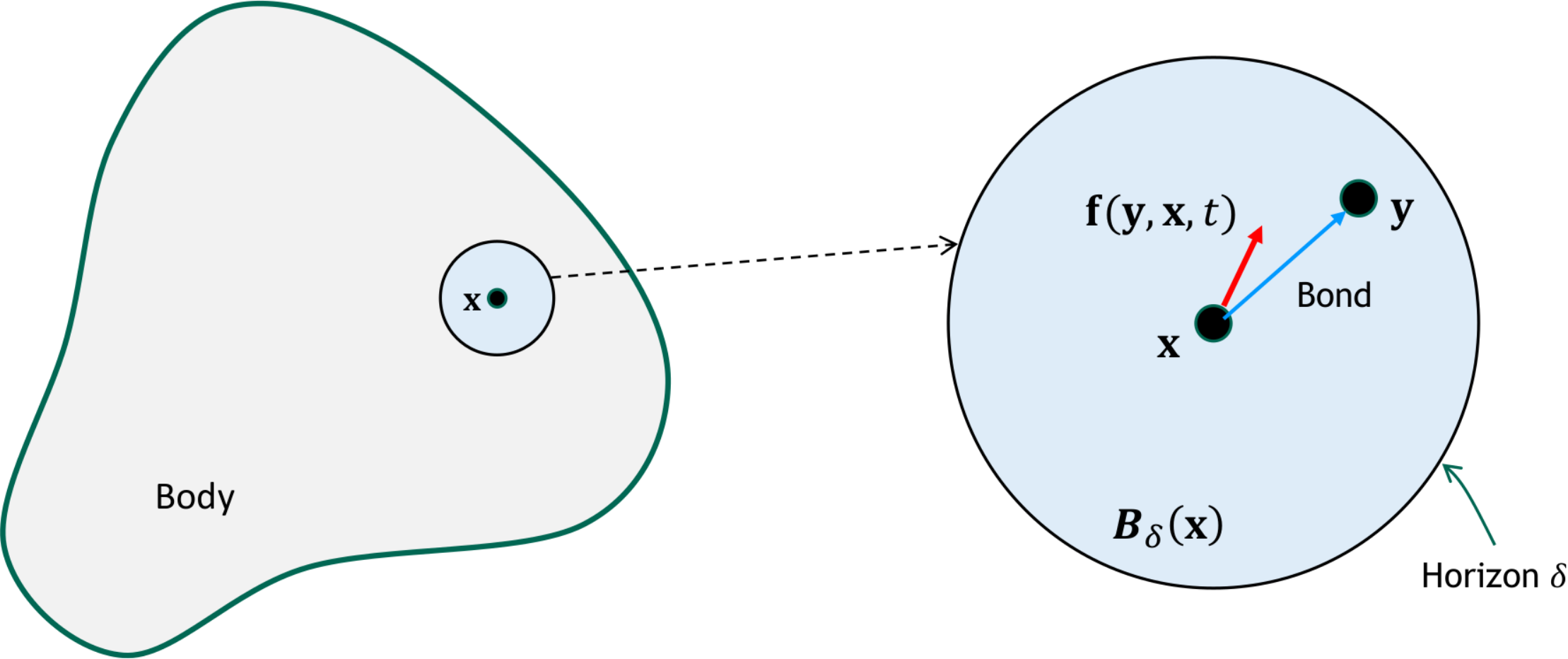}
\caption{Left: the family of a point $\xb$ in a peridynamic body.
Right: typical bond and bond force vector.}
\label{fig-potato}
\end{figure}

In the previous section, a coarse-grained continuum momentum balance was derived, given by \eqref{eqn-eom}.
This momentum balance has a fundamentally nonlocal character, since the {\emph{pairwise bond force densities}} given by \eqref{eqn-fdef} can be nonzero whenever the material points in the continuum $\xb$ and $\yb$ are separated by a finite distance up to the horizon $\delta$ (Figure~\ref{fig-potato}).
This form of the momentum balance is known as the {\emph{peridynamic}} equation of motion \cite{silling_2000}.
In peridynamics, each $\xb$ interacts through bond forces with other material points $\yb$ within a neighborhood with radius $\delta$ known as the {\emph{family}} of $\xb$, denoted by ${B_\delta(\xb)}$.
The equation of motion for material point $\xb$ is then 
\begin{equation}
  \rho(\xb)\ddot\ub(\xb,t)= \int_{{B_\delta(\xb)}}\fb(\yb,\xb,t)\;d\yb+\bb(\xb,t).
    \label{eqn-pdeom}
\end{equation}
A {\emph{material model}} in peridynamics supplies values of $\fb(\yb,\xb,t)$ in terms of the deformations of the families
of $\xb$ and $\yb$ and any other relevant variables such as temperature.
In general, material models in peridynamics are specified using operators called {\emph{states}} that are nonlocal analogues of second order tensors \cite{silling_2007}.
Many material models have been developed for peridynamics, and any material model from the local theory can be translated into peridynamic form \cite{silling_2010}.
The most widely used capability that peridynamics offers that is not available in the local theory is the direct modeling of fracture within the basic field equations.
Peridynamics can model fracture because the equation of motion \eqref{eqn-pdeom} is an integro-differential equation that does not involve the partial derivatives of displacement with respect to position.
However, the present paper concerns only small deformations in the linear regime of material response and does not address fracture.
The extension of the methods described here to determine a linear peridynamic material model to the nonlinear regime, including fracture, is under investigation in separate work.
The remainder of this paper deals with a specific material model described in the next section. Note that even though MD displacements are dynamic, we smooth them in time as described in \eqref{eqn:smoothu} so that the time-space smoothed MD data can be described by the static counterpart of the nonlocal equation \eqref{eqn-eom},
\begin{equation}
-\int_{B_\delta(\xb)} \fb(\yb,\xb,\ub)\,d\yb =\bb(\xb),
    \label{eqn-pdeom_static}
\end{equation}
where $\fb$ is to be determined (see the following section) and where we introduced the nonlocal interaction region $B_\delta(\xb)$. The {\emph{horizon}} $\delta$ determines the extent of the nonlocal interactions. Although, according to Section \ref{sec:smoothing}, $\delta$ could be determined by the cutoff distance $d$ associated with $\Fb$ and the radius of the smoothing function $R$, in the following discussion it is treated as a learned parameter (Section \ref{sec:learning-results}). This approach allows for a finite value of $\delta$ to be obtained even if $d=\infty$, as would be the case with the Lennard-Jones potential.

\subsection{The Linear Peridynamic Solid (LPS) Model}

The main application considered in this work is the simulation of displacements in single-layered graphene. The graphene sheet is treated using a two-dimensional nonlocal model under the assumption of plane stress, which is appropriate for a thin sheet. The pairwise bond force density $\fb$ is determined using the state-based linear peridynamic solid (LPS) model. The LPS model is a prototypical state-based model appropriate for isotropic elastic materials. It may be regarded as a nonlocal generalization of the local model for an isotropic solid, which contains contributions from shear and dilatation. The LPS model has advantages over the previously developed bond-based peridynamic models in that it is not restricted to a Poisson’s ratio of 1/4. The LPS model has known well-posedness properties under certain assumptions  \cite{emmrich2007well}. This section summarizes the mathematical formulation for the LPS model and illustrates a meshfree discretization \cite{trask2019asymptotically,yu2021asymptotically,fan2021asymptotically,you2019asymptotically,you2020asymptotically,foss2021convergence}.

Consider a 2D body occupying the domain $\Omega\subset\mathbb{R}^2$, and let $\theta$ be the nonlocal dilatation, generalizing the local divergence of the displacement. Let $K(r)$ be the influence function \cite{seleson2011role} which modulates nonlocal effects within a peridynamic model. In this work, we assume $K$ to be a radial function compactly supported on the $\delta$-ball $B_\delta(\mathbf{x})$ with $\alpha$-th order singularity:
\begin{equation}\label{eqn:require_ga}
K(\xb,\yb):=K(|\xb-\yb|)=\frac{P(|\xb-\yb|)}{|\xb-\yb|^\alpha}
\end{equation}
where $P(r)$ is a bounded function in $[0,\delta]$. The momentum balance and nonlocal dilatation are given by
\begin{equation}\label{eq:nonlocElasticity}
\begin{aligned}
    \mcL_K \mathbf{u}(\xb):=&-\frac{C_1}{m(\delta)}  \int_{B_\delta (\mathbf{x})} \left(\lambda- \mu\right) K(\left|\mathbf{y}-\mathbf{x}\right|) \left(\mathbf{y}-\mathbf{x} \right)\left(\theta(\mathbf{x}) + \theta(\mathbf{y}) \right) d\mathbf{y}\\
  &-  \frac{C_2}{m(\delta)}\int_{B_\delta (\mathbf{x})} \mu K(\left|\mathbf{y}-\mathbf{x}\right|)\frac{\left(\mathbf{y}-\mathbf{x}\right)\otimes\left(\mathbf{y}-\mathbf{x}\right)}{\left|\mathbf{y}-\mathbf{x}\right|^2}  \left(\mathbf{u}(\mathbf{y}) - \mathbf{u}(\mathbf{x}) \right) d\mathbf{y} = \bb(\xb),
  \end{aligned}
\end{equation}
\begin{equation}\label{eqn:oritheta}
\theta(\mathbf{x}):=\dfrac{2}{m(\delta)}\int_{B_\delta (\mathbf{x})} K(\left|\mathbf{y}-\mathbf{x}\right|) (\mathbf{y}-\mathbf{x})\cdot \left(\mathbf{u}(\mathbf{y}) - \mathbf{u}(\mathbf{x}) \right)d\mathbf{y},
\end{equation}
where $\mathbf{u}\in\mathbb{R}^2$ denotes the displacement, $\bb\in\mathbb{R}^2$ the prescribed body force density, and $m(\delta)$ the weighted volume. In the present notation, the nonlocal operator $\mcL_K[\ub](\xb)$ in \eqref{eq:nonlocElasticity} has the subscript $K$ to emphasize its dependence on the influence function $K$. This operator corresponds to the integral term $-\int \fb(\yb,\xb,\ub)d\yb$ in \eqref{eqn-eom}, or, equivalently, \eqref{eqn-pdeom_static}.

{Given a forcing term $\bb$, in order to guarantee the existence of a unique solution $\ub$, ``nonlocal boundary conditions'', or volume constraints, must be prescribed on an appropriate {\it interaction} domain $\Omega_I$, so that the LPS problem becomes
\begin{equation}\label{coarsegrained}
\left\{
\begin{aligned}
\mathcal{L}_K[\ub](\xb) &= \bb(\xb) \quad \xb \in \Omega,\\
\mathcal{B}_{I} \ub(\xb) &= \mathbf{q}(\xb) \quad  \xb \in \Omega_I.
\end{aligned}
\right.
\end{equation}
Here, $\mathcal{B}_I$ is a nonlocal interaction operator specifying a volume constraint. In this work, without loss of generality, we consider the Dirichlet condition $\mathcal{B}_I = \mcI$, where $\mcI$ is the identity operator. Other types of conditions, e.g., Neumann \cite{DEliaNeumann2020,DEliaBC2021,you2019asymptotically,yu2021asymptotically}, Robin \cite{yu2018partitioned,you2020asymptotically} or periodic \cite{You2020Regression}, are also compatible with our learning algorithm.}

\paragraph{A meshfree discretization of the LPS model}

{Given a collection of material points $\mcX = \{ \mathbf{x}_i \}_{i=1,2,\dots,N_p}$, we numerically evaluate $\mathcal{L}_K(\ub)$ by employing the meshfree, particle discretization introduced in \cite{yu2021asymptotically}, which features ease of implementation and efficiency.} At each material point $\xb_i$, we adopt the following quadrature rule to approximate the integral in $\mcL_K$ in \eqref{eq:nonlocElasticity}, which we now denote by $\mcL_K^h$.
\begin{equation}\label{eq:discrete}
\begin{aligned}
    \mcL_K^h \mathbf{u}(\xb_i):=-\frac{C_1}{m_i(\delta)}  \sum_{ \mathbf{x}_j \in B_\delta (\mathbf{x_i})} \left(\lambda- \mu\right) K_{ij} \left(\mathbf{x}_j-\mathbf{x}_i \right)\left(\theta^h(\mathbf{x}_i) + \theta^h(\mathbf{x}_j) \right) W_{j,i}\\
  -  \frac{C_2}{m_i(\delta)}\sum_{\mathbf{x}_j \in B_\delta (\mathbf{x}_i)} \mu K_{ij}\frac{\left(\mathbf{x}_j-\mathbf{x}_i\right)\otimes\left(\mathbf{x}_j-\mathbf{x}_i\right)}{\left|\mathbf{x}_j-\mathbf{x}_i \right|^2}  \left(\mathbf{u}(\xb_j) - \mathbf{u}(\xb_i) \right) W_{j,i} = \bb(\xb_i),
\end{aligned}
\end{equation}
\begin{equation}\label{eqn:disctheta}
\theta^h(\mathbf{x}_i):=\dfrac{2}{m_i(\delta)} \sum_{ \mathbf{x}_j \in B_\delta (\mathbf{x}_i)} K_{ij} (\mathbf{x}_j-\mathbf{x}_i)\cdot \left(\mathbf{u}(\xb_j) - \mathbf{u}(\xb_i) \right)W_{j,i},
\end{equation}
where $K_{ij} := K(\mathbf{x}_j,\mathbf{x}_i)$ and $m_i(\delta):=\underset{\mathbf{x}_j \in B_\delta (\mathbf{x}_i)}{\sum}K_{ij}\verti{\mathbf{x}_j-\mathbf{x}_i}^2 W_{j,i}$. The quadrature weights $W_{j,i}$ are obtained for material points on different subdomains $B_\delta(\xb_i)$, from the following optimization problem
\begin{align}\label{eq:quadQP}
  \underset{\left\{\omega_{j,i}\right\}}{\text{argmin}} \sum_{\xb_j \in \mcX_h\cap B_\delta(\xb_i)\backslash\{\xb_i\}} \!\!W_{j,i}^2 \quad
  \text{s. t.}, \;
  \sum_{\xb_j\in B_\delta(\xb_i)}\!\!q(\xb_i,\xb_j)W_{j,i} = \int_{B_\delta(\xb_i)} q(\xb_i,\yb)d\yb \quad \forall \,q \in \bm{V},
\end{align}
where $\bm{V}$ denotes the space of functions which should be integrated exactly. Following \cite{yu2021asymptotically}, in this work we take $\bm{V}:=\left\{q(\yb)= \frac{p(\yb)}{|\yb-\xb|^3} \,|\, p \in \bm{P}_5(\mathbb{R}^2) \text{ such that } \int_{B_\delta(\xb)} q(\yb) d \yb < \infty \right\}$ and $\bm{P}_5(\mathbb{R}^2)$ denotes the space of quintic polynomials . 

Although the developed learning approach as well as the meshfree quadrature rule can be applied to the general collection of material points $\mcX$, in this work we consider the uniform Cartesian grid for simplicity:
$$
\mcX_h:=\{(p_1h,p_2h)|\bm{p}=(p_1,p_2)\in\mathbb{Z}^2\}\cap(\Omega\cap\Omega_I),
$$ 
where $h$ is the spatial grid size. As we discuss in the next section and in Section \ref{sec:generalization}, different grids $\mcX_h$ can be used for training and validation sample collection.

\begin{remark}
Using the same arguments as in \cite{fan2021asymptotically}, it can be seen that the chosen quadrature rule provides a consistent approximation of $\mathcal{L}_K(\ub)$ when $\alpha<3$. Therefore, in the learning algorithm,  we require the fractional order $\alpha$ to be bounded by $3$, and we note that this requirement may be further relaxed by considering other discretization methods.
\end{remark}

\section{Operator regression for the LPS model}\label{sec:regression}

In this section we illustrate how to extend the data-driven approach developed in \cite{You2020Regression,You2021}, known as {\emph{nonlocal operator regression}}, to the LPS model. Section \ref{sec:operator} first describes the regression algorithm under the assumption that coarse-grained MD displacements are available at any material point. Then, Section \ref{sec:infsup} introduces solvability constraints that guarantee that the optimal nonlocal model is well-posed by construction\footnote{The solvability conditions derived in \cite{You2020Regression} for 1D nonlocal diffusion problems are not applicable to the more complex LPS model considered in the current work.}. Lastly, Section \ref{sec:workflow} summarizes the complete workflow of the data-driven nonlocal operator regression algorithm, which is the process of going from high fidelity MD simulations to data-driven optimal kernels via coarse graining and operator regression.

\subsection{Operator regression algorithm}\label{sec:operator}

The foundation of the nonlocal operator regression algorithm is the fact that a coarse-grained displacement $\ub$ follows a nonlocal evolution law of the form  \eqref{coarsegrained}. For this nonlocal model, we seek to identify an optimal constitutive relation on the basis of MD data sets.

Let $\{\ub^s(\xb_{i,s}),\bb^s(\xb_{i,s})\}$, $s=1,\cdots,S$, be given pairs of displacement and body force fields available at $\xb_{i,s}\in\mcX^s$, and let $\mathcal{L}_K$ be the LPS operator defined in \eqref{eq:nonlocElasticity} parametrized by the material properties $\lambda$ and $\mu$ and by the influence function $K$. We aim to learn an optimal nonlocal operator $\mcL_K$. This optimal operator consists of the influence function $K$, which may be sign-changing, and parameters $\lambda$ and $\mu$, such that the action of $\mathcal{L}_K$ most closely maps $\ub^s(\xb)$ to $\bb^s(\xb)$ for all $s$. Formally, the optimal influence function and parameters, $(\lambda^*,\mu^*,K^*)$, are the solution of the following optimization problem:
\begin{equation}\label{abstractOptProblem}
(K^*,\lambda^*,\mu^*) = \underset{\lambda,\mu,K}{\text{argmin}}{\frac{1}{S}} \sum_{s=1}^S \big\| \mathcal{L}^h_K [\ub^s](\xb_{i,s}) - \bb^s(\xb_{i,s}) \big\|^2_{\ell_2(\mcX^s)}.
\end{equation}
To increase the flexibility of the algorithm, each sample can be available on different point sets $\mcX^s$.

The influence function $K(\xb,\yb)$ will now be parameterized. Following \cite{You2020Regression}, assume that $K$ has the form of \eqref{eqn:require_ga}, and represent its numerator $P$ as a linear combination of Bernstein polynomials evaluated at $|\xb-\yb|$:
\begin{equation}\label{eqn:K}
    K(\mathbf{x},\mathbf{y}) = \sum_{k=0}^{M} \frac{D_k}{|\mathbf{x}-\mathbf{y}|^\alpha}B_{k,M}\bigg (\frac{|\mathbf{x}-\mathbf{y}|}{\delta} \bigg).     
\end{equation}
Here the Bernstein polynomials are defined as 
\begin{equation}
    B_{k,M}(r) = \begin{pmatrix}
    M \\
    k \\
    \end{pmatrix}
    r^k(1-r)^{M-k}, \quad \text{ for } 0 \leq r \leq 1.
\end{equation}
To allow the learning of nonlocal models whose kernels may be partially negative, we allow $D_k\in\mbR$, for all $k$. This generality, however, might compromise the well-posedness of the resulting optimal model, since known well-posedness results on LPS models only apply to positive kernels \cite{mengesha14Navier}. To guarantee that the LPS model associated with $(\lambda^*,\mu^*,K^*)$ is solvable by construction, we embed in our algorithm sufficient well-posedness conditions for the discretized operator; these are described in detail in the next section.

The formulation of the constrained optimization problem is as follows. Given a collection of training samples $\{\ub^s(\xb_{i,s}),\bb^s(\xb_{i,s})\}$, $s=1,\cdots,S$, we seek to learn the parameters $\lambda$ and $\mu$, the Bernstein polynomial coefficients $\Db=[D_0,\cdots,D_M]\in\real^{M+1}$, the order $\alpha$, the horizon $\delta$, and the polynomial order $M$ by minimizing the mean square loss (MSL) of the LPS equation:
\begin{equation}\label{eq:true_problem} 
\left\{
\begin{aligned}
&(\lambda^*,\mu^*,\Db^*,\alpha^*,\delta^*,M^*)=\underset{\lambda,\mu,\Db,\alpha,\delta,M}{\text{argmin }}\; \frac{1}{S} \sum_{s=1}^S ||\mcL_{K}^h\ub^s(\xb_{i,s}) - \bb^s(\xb_{i,s}) ||^2_{\ell_2(\mcX^s)} \\
&\text{s.t. solvability constraints.}
\end{aligned}
\right.
\end{equation}

\subsection{Solvability Constraints for the Discretized LPS Model}\label{sec:infsup}

When the influence function $K$ as described in \eqref{eqn:require_ga} is nonnegative and $\alpha< 4$, the LPS model is well-posed, as shown, for example, in \cite{mengesha14Navier}. However, several works have indicated the practical need for sign-changing kernels \cite{weckner2011determination,mengesha2013analysis,xu2020deriving,xu2021learning,You2021}. While it is unclear whether multiscale physics inherently leads to sign-changing kernels or if equally descriptive positive kernels could be derived, in \cite{You2020Regression} the authors found that 
allowing for sign-changing kernels provides a significant increase in accuracy when modeling high-frequency material response. Therefore, in this work we seek a well-posed LPS model with possibly sign-changing influence functions $K$. As there is no available theory on sufficient conditions for the well-posedness of LPS models with sign-changing $K$, we impose well-posedness conditions on the discretized system directly; as a result, well-posedness of the learnt model is guaranteed for the discretization method used during training.

An inequality constraint for the well-posedness of the meshfree discretization approach \eqref{eq:discrete}-\eqref{eqn:disctheta} that allows for sign-changing influence functions will now be derived. For simplicity of analysis, and without loss of generality, assume homogeneous Dirichlet-type boundary conditions: $\ub(\xb)=0$ in $\Omega_I$.

For the derivation of the solvability constraints that will be employed in our algorithm to ensure the well-posedness of the discretized LPS model, first write the discretized LPS model \eqref{eq:discrete}-\eqref{eqn:disctheta} as the following linear system:
\begin{equation}\label{eqn:discsystem}
    \begin{pmatrix}
    \mu \Gamma & (\Phi)^{\mathrm{t}} \\
    \Phi & -\frac{1}{\lambda-\mu}I  
    \end{pmatrix}
    \begin{pmatrix}
    \mathbf{U} \\
    \bm{\Theta}\\ 
    \end{pmatrix}
    =\begin{pmatrix}
    -\mathbf{B} \\
    \mathbf{0}\\
    \end{pmatrix}.
\end{equation}
Here, $\mathbf{U}\in \real^{2N_p}$ and $\frac{1}{\lambda-\mu}\bm{\Theta}\in\real^{N_p}$ are the vectors of the degrees of freedom (DOFs) of the displacement $\ub$ and the nonlocal dilatation $\theta$:
$$
\mathbf{U}=[(\ub(\xb_1))^t,\cdots,(\ub(\xb_{N_p}))^t]^t, \quad \bm{\Theta}=[(\lambda-\mu)\theta(\xb_1),\cdots,(\lambda-\mu)\theta(\xb_{N_p})]^t.
$$
$\mathbf{B}$ is the vector of DOFs of the body load and has the same length and ordering of indices as $\mathbf{U}$. $I$ is an $N_p\!\times\!N_p$ identity matrix, and $\Gamma$ and $\Phi$ are the matrices that correspond to the deviatoric and dilatation contributions of the deformation:
$$\Gamma \mathbf{U}=[(\fb_{dev}(\xb_1))^t,\cdots,(\fb_{dev}(\xb_{N_p}))^t]^t,\quad \Phi^t \mathbf{U}=[f_{dil}(\xb_1),\cdots,f_{dil}(\xb_{N_p})]^t,$$
where
\begin{align*}
    \fb_{dev}(\xb_i) &=  -\frac{C_2}{m_i(\delta)}\sum_{\mathbf{x}_j \in B_\delta (\mathbf{x}_i)}  K_{ij}\frac{\left(\mathbf{x}_j-\mathbf{x}_i\right)\otimes\left(\mathbf{x}_j-\mathbf{x}_i\right)}{\left|\mathbf{x}_j-\mathbf{x}_i \right|^2}  \left(\mathbf{u}(\xb_j) - \mathbf{u}(\xb_i) \right) W_{j,i}\\
    f_{dil}(\xb_i) &=\dfrac{2}{m_i(\delta)} \sum_{ \mathbf{x}_j \in B_\delta (\mathbf{x}_i)} K_{ij} (\mathbf{x}_j-\mathbf{x}_i)\cdot \left(\mathbf{u}(\xb_j) - \mathbf{u}(\xb_i) \right)W_{j,i}.
\end{align*}

In what follows, for each vector $\Vb\in\real^{2N_p}$, write $\Vb=[\vb_1^t,\cdots,\vb_{N_p}^t]^t$ with each $\vb_i\in\real^2$. $C$ denotes a generic constant.
The following theorem provides sufficient conditions that guarantee the solvability of \eqref{eqn:discsystem}. Let the energy ``norm'' be defined as $\vertii{\Vb}^2_{E}:=\Vb^t \Gamma\Vb$ for all $\Vb\in\mathbb E\setminus{\bf 0}$, where $\mathbb{E}$ denotes the quotient space of $\real^{2N_p}$ by the discrete space of infinitesimally rigid displacements:
$$
\Pi_\mcX=\{[(\mathbb{Q}\xb_1+\db)^t,\cdots,(\mathbb{Q}\xb_{N_p}+\db)^t]^t,\;\mathbb{Q}\in\real^{2\times 2},\;\mathbb{Q}^T=-\mathbb{Q},\;\db\in\real^2\}.$$
Note that $\|\cdot\|_E$ is indeed a norm only when certain conditions, reported in the following theorem, are satisfied.

\begin{thm}[Sufficient Conditions for Solvability]\label{lem:suf}
The discretized LPS formulation \eqref{eqn:discsystem} is solvable for any values of $\lambda+\mu>0$ and $\mu>0$, provided that the following conditions hold:
\begin{align}
&\text{i. Discrete Continuity and Coercivity: }
\exists\, \underline{C}_\Gamma>0,\, \overline{C}_\Gamma<\infty \;\text{ s.t.,}
\\[2mm]
&\quad\vertii{\Vb}^2_{E}\geq \underline{C}_\Gamma \vertii{\Vb}^2_{\ell^2}\;\; \text{and} \;\; \vertii{\Vb}^2_{E}\leq \overline{C}_\Gamma \vertii{\Vb}^2_{\ell^2} \quad\forall\,\Vb\in \mathbb{E}\backslash \{\bm{0}\}\label{eqn:ellip}
\\[3mm]
&\text{ii. Discrete Inf-Sup: }\exists\, C_\Phi>0,\text{ s.t.,}\underset{\Pb\in\real^{N_p
}\backslash\{\bm{0}\}}{\inf}\;\underset{\Vb\in\mathbb{E}\backslash\{\bm{0}\}}{\sup}\frac{\Vb^t \Phi^t\Pb}{\vertii{\Vb}_{E}\vertii{\Pb}_{\ell^2}} \geq C_\Phi,\label{eqn:infsup}
\\[3mm]
&\text{iii. Discrete, generalized Cauchy-Schwarz: }\vertii{\Vb}^2_{E}\geq  2\vertii{\Phi\Vb}^2_{\ell^2},\;\forall\,\Vb\in \mathbb{E}.\label{eqn:ebound}
\end{align}
\end{thm}
\begin{remark}
In the continuous case with $K\geq0$, the property \eqref{eqn:ebound} is an immediate result from the Cauchy-Schwarz inequality, see, for example, \cite{mengesha14Navier}. This property yields the equivalence of the semi-norm from the deviatoric part of the deformation and the full strain energy.
\end{remark}

\begin{proof}
Inequality \eqref{eqn:ellip} implies that the energy norm is a norm in $\mathbb E\setminus\{\bf 0\}$. We consider two scenarios: $\lambda-\mu\geq0$ and $\lambda-\mu<0$.

\noindent \underline{Case 1}: $\lambda-\mu\geq0$. The symmetry property of the Cartesian grids implies that $m_i(\delta)=m_j(\delta):=m$ and $W_{j,i}=W_{i,j}$ for all $i,j\in\{1,\cdots,N_p\}$. By taking the inner product of \eqref{eqn:discsystem} with $(\Vb^t,\bm{\Xi}^t)$, component-wise, we reformulate the system as a general mixed formulation: find
$\Ub\in\real^{2N_p}$ and $\bm{\Theta} \in \real^{N_p}$ such that
\begin{align*}
    &a(\Ub,\Vb) + b(\Vb,\bm{\Theta}) = (-\mathbf{B},\Vb), \quad &\forall\, \Vb \in \real^{2N_p}\\
    &b(\Ub,\bm{\Xi}) - c(\bm{\Theta},\bm{\Xi}) = 0, \quad &\forall\, \bm{\Xi} \in \real^{N_p}.
\end{align*}
Here $(\cdot,\cdot)$ denotes the inner product, $a(\Ub,\Vb):=\mu \Vb^t\Gamma\Ub$, $b(\Vb,\bm{\Xi}):=\bm{\Xi}^t\Phi\Vb$, and $c(\bm{\Theta},\bm{\Xi}):=\frac{1}{\lambda-\mu}\bm{\Theta}^t\bm{\Xi}$. Firstly, note that when \eqref{eqn:ellip} is satisfied, the symmetric bilinear form $a(\cdot,\cdot)$ is continuous and coercive.
Similarly,
\begin{align*}
b(\Vb,\bm{\Xi})=&-\frac{C_1}{m}  \sum_{i=1}^{N_p} \left(\sum_{ \mathbf{x}_j \in B_\delta (\mathbf{x_i})} \left(\lambda- \mu\right) K_{ij} \vb_i^t\left(\mathbf{x}_j-\mathbf{x}_i \right) \left(\xi_i+\xi_j \right) W_{j,i}\right)\\
=&\frac{C_1}{2m}  \sum_{i=1}^{N_p} \left(\sum_{ \mathbf{x}_j \in B_\delta (\mathbf{x_i})} \left(\lambda- \mu\right) K_{ij} (\vb_j-\vb_i)^t\left(\mathbf{x}_j-\mathbf{x}_i \right) \left(\xi_i+\xi_j \right) W_{j,i}\right)\\
\leq&C\vertii{\Vb}_{\ell^2}\vertii{\bm{\Xi}}_{\ell^2}\leq C\vertii{\Vb}_{E}\vertii{\bm{\Xi}}_{\ell^2}
\end{align*}
so that $b(\cdot,\cdot)$ is also a continuous bilinear form. By combining \eqref{eqn:ellip} and \eqref{eqn:infsup} with the fact that $c(\bm{\Theta},\bm{\Xi})=\frac{1}{\lambda-\mu}\bm{\Theta}^t\bm{\Xi}\leq C\vertii{\bm{\Theta}}_{\ell^2}\vertii{\bm{\Xi}}_{\ell^2}$ and $c(\bm{\Xi},\bm{\Xi})=\frac{1}{\lambda-\mu}\vertii{\bm{\Xi}}^2_{\ell^2}\geq 0$, the well-posedness of \eqref{eqn:discsystem} follows using the same arguments of \cite[Section~2.2]{Bathe:6}.

\noindent \underline{Case 2}: $\lambda-\mu< 0$. In this case $c(\bm{\Xi},\bm{\Xi})$ is not coercive with respect to the $\ell^2$ norm and therefore the theory in \cite{Bathe:6} does not apply. By reducing the discrete system \eqref{eqn:discsystem} to $((\lambda-\mu)\Phi^t\Phi+\mu\Gamma)\Ub=-\mathbf{B}$, we note that $(\lambda-\mu)\Phi^t\Phi+\mu\Gamma$ is not solvable, or, equivalently, non-invertible, if and only if there exists $\Vb\in\mathbb{E}\backslash\{\bm{0}\}$ such that $((\lambda-\mu)\Phi^t\Phi+\mu\Gamma)\Vb=\bm{0}$. We prove that this is not possible by showing that
$$\Vb^t((\lambda-\mu)\Phi^t\Phi+\mu\Gamma)\Vb=0\Leftrightarrow \Vb=\bm{0}.$$
We split the proof in two parts: $\lambda\geq 0$ and $\lambda<0$, respectively. First, for $\lambda\geq0$, 
\eqref{eqn:ellip} and \eqref{eqn:ebound} yield
\begin{align*}
0=&\Vb^t((\lambda-\mu)\Phi^t\Phi+\mu\Gamma)\Vb= \lambda\Vb^t\Phi^t\Phi\Vb+\mu(\vertii{\Vb}_E^2-\vertii{\Phi\Vb}_{\ell^2}^2)\\[2mm]
\geq &\lambda\Vb^t\Phi^t\Phi\Vb+\frac{\mu}{2}\vertii{\Vb}_E^2\geq \frac{\mu}{2}\vertii{\Vb}_{\ell^2}^2.
\end{align*}
Hence $\Vb=\bm{0}$, which contradicts our assumption. Similarly, when $\lambda<0$, we assume that $\vertii{\Vb}_E^2>0$. From the assumptions $\lambda-\mu<0$ and $\lambda+\mu>0$ we have
\begin{align*}
0=&\Vb^t((\lambda-\mu)\Phi^t\Phi+\mu\Gamma)\Vb\geq \frac{\lambda-\mu}{2}\Vb^t\Gamma\Vb+\mu\vertii{\Vb}_E^2= 
\frac{\lambda+\mu}{2}\vertii{\Vb}_E^2,
\end{align*}
which, again, implies $\Vb=\bm{0}$, contradicting our assumption. Therefore, $(\lambda-\mu)\Phi^t\Phi+\mu\Gamma$ is invertible and \eqref{eqn:discsystem} is solvable.
\end{proof}

\begin{remark}
When the influence function $K$ is nonnegative, the continuous LPS model satisfies the ellipticity condition and the inf-sup condition. Moreover, the energy density associated with the dilatation part is bounded by energy density associated with the deviatoric part of the deformation, as shown in \cite{mengesha14Navier}. These facts, intuitively, support our well-posedness result, where the constraints \eqref{eqn:ellip}, \eqref{eqn:infsup} and \eqref{eqn:ebound} are equivalent to requiring that the discretized LPS model associated with a sign-changing $K$ satisfies the same three properties as its nonnegative-kernel counterpart.
\end{remark}

Therefore, we augment our learning problem \eqref{eq:true_problem} with the inequality (solvability) constraints corresponding to \eqref{eqn:ellip}, \eqref{eqn:infsup} and \eqref{eqn:ebound}. The constants $\underline{C}_\Gamma$, $\overline{C}_\Gamma$, and $C_\Phi$ are only related to the influence function $K$ and are independent of the shear and Lam\'e parameters $\lambda$ and $\mu$. We calculate the inf-sup constant $C_\Phi$ in \eqref{eqn:infsup} by solving an eigenvalue problem, as indicated by the following theorem.
\begin{thm}
The inf-sup constant in \eqref{eqn:infsup} in Theorem \ref{lem:suf} can be expressed as
$$C_\Phi=\Lambda_{min}(\Phi\Gamma^{\dagger}\Phi^t)$$
where, for a given a square matrix $M$, $M^\dagger$ denotes its pseudoinverse and $\Lambda_{min}(M)$ denotes its smallest nonzero eigenvalue.
\end{thm}

\begin{proof}
The proof can be obtained following the same arguments as in \cite[Proposition~3.1]{brezzi2012mixed}. 
\end{proof}

For any given $K$ as in \eqref{eqn:require_ga} and a fixed discretization method, the largest eigenvalue of $\Gamma$ is bounded by construction. Therefore $\overline C_\Gamma$ is finite and, in practice, \eqref{eqn:ellip}, \eqref{eqn:infsup} and \eqref{eqn:ebound} can be imposed as:
\begin{align}\label{eq:sufficient-constraints}
&\Lambda_{min}(\Gamma)\geq \zeta,\\
&\Lambda_{min}(\Phi\Gamma^{\dagger}\Phi^t)\geq \zeta,\\
&\Lambda_{min}(\Gamma-2\Phi^t\Phi)\geq 0,
\end{align}
where $\zeta>0$ is a given, small number. 

We can now state the solvability-constrained optimization problem. Rename the matrices $\Gamma$ and $\Phi$ in \eqref{eqn:discsystem} as $\Gamma_{(\alpha,\Db,\delta,M)}$ and $\Phi_{(\alpha,\Db,\delta,M)}$ indicating that they are parameterized with the Bernstein polynomial coefficients $\Db$, the fractional order $\alpha$, the horizon $\delta$, and the highest Bernstein polynomial order $M$. Given the tolerance parameter $\zeta>0$, the learning problem is stated as
\begin{equation}\label{eq:true_problem_full} 
\left\{
\begin{aligned}
(\lambda^*,\mu^*,\Db^*,\alpha^*,\delta^*,M^*)&=\underset{\lambda,\mu,\Db,\alpha,\delta,M}{\text{argmin }} \; \frac{1}{S} \sum_{s=1}^S ||\mcL_{K}^h\ub^s(\xb_{i,s}) - \bb^s(\xb_{i,s}) ||^2_{\ell_2(\mcX^s)} \\
\text{subject to:}\quad &\lambda+\mu>0,\,\mu>0, \, \alpha<3, \,\Lambda_{min}(\Gamma_{(\alpha,\Db,\delta,M)})\geq \zeta,\\
\quad &\Lambda_{min}(\Phi_{(\alpha,\Db,\delta,M)}\Gamma^{\dagger}_{(\alpha,\Db,\delta,M)}\Phi^t_{(\alpha,\Db,\delta,M)})\geq \zeta, \\
\quad &\Lambda_{min}(\Gamma_{(\alpha,\Db,\delta,M)}-2\Phi^t_{(\alpha,\Db,\delta,M)}\Phi_{(\alpha,\Db,\delta,M)})\geq 0.
\end{aligned}
\right.
\end{equation}


\begin{remark}
The constraints in \eqref{eq:sufficient-constraints} are sufficient to guarantee that the model associated with the optimal influence function $K$ is well-posed only when discretized with the same technique and resolution utilized during training. Being only sufficient, these conditions may yield an optimal influence function whose associated model is still well-posed when discretized with different schemes or resolutions. More discussion and numerical tests on this topic can be found in Section \ref{sec:generalization}.
\end{remark}

\subsection{Algorithm and Workflow} \label{sec:workflow}

\begin{algorithm}
\caption{Two-stage strategy to solve \eqref{eq:true_problem_full}  for $(\lambda^*,\mu^*,\alpha^*,\mathbf{D}^*)$.}\label{alg:augmented_lagrangian}
\begin{algorithmic}[1]
\State With fixed $\delta$ and $M$, initialize $\lambda$, $\mu$, $\alpha$, $D_k^{(0)} \sim \mathcal{U}\left(0,1\right)$,where $\mathcal{U}\left(a,b\right)$ denotes the uniform distribution on $\left(a,b\right)$.
\State
Obtain $(\lambda^{pre},\mu^{pre},\alpha^{pre},\Db^{pre})$ as a local minimum of $L^{pre}(\lambda,\mu,\alpha,\Db)$, using the Adam optimizer while
updating $\lambda \leftarrow \text{ReLU}(\lambda)$, $\mu \leftarrow \mu$, $\alpha \leftarrow 3-\text{ReLU}(3-\alpha)$ and $\mathbf{D} \leftarrow \text{ReLU}(\mathbf{D})$ after each step of gradient descent. 
\State
Initialize $(\lambda^{(0)},\mu^{(0)},\alpha^{(0)},\Db^{(0)})=(\lambda^{pre},\mu^{pre},\alpha^{pre},\Db^{pre})$ and $\varpi^{(0)}_1=\varpi^{(0)}_2=\varpi^{(0)}_3=1$.
\State Set $\text{\texttt{STEP\_{MAX}}}=100$, $\phi_1=\phi_2=\phi_3=0$, $\psi=1$, $r_1=5$, $r_2=1/4$, $\epsilon=10^{-8}$.

\While{$j \leq \text{\texttt{STEP\_{MAX}}}$:} \Comment{Perform Augmented Lagrangian Algorithm}
\State\label{step_solve_optimization}
Solve the unconstrained optimization problem
\begin{align*}
    &(\lambda^{(j)},\mu^{(j)},\alpha^{(j)},\Db^{(j)},\bm{\varpi}^{(j)})=\underset{\lambda,\mu,\alpha,\mathbf{D}, \bm{\varpi}}{\text{argmin }} 
    L^{corr}(\lambda,\mu,\alpha,\mathbf{D}, \bm{\varpi}).
\end{align*}
\If{$H_p(\alpha^{(j)},\mathbf{D}^{(j)},\bm{\varpi}^{(j)})\leq \epsilon$ for all $p\in\{1,2,3\}$,}
\State
Stop.
\Else
\If{$\exists \,p\in\{1,2,3\}$ s.t. $H_p(\alpha^{(j)},\mathbf{D}^{(j)},\bm{\varpi}^{(j)})\geq r_2H_p(\alpha^{(j-1)},\mathbf{D}^{(j-1)},\bm{\varpi}^{(j-1)})$}
\State
Update penalty $\psi\leftarrow r_1\psi$.
\If{$\psi\geq 10^{20}$}
\State
Stop.
\EndIf
\Else
\State
Update Lagrange multiplier $\phi_p \leftarrow \phi_p+\psi H_p(\alpha^{(j)},\mathbf{D}^{(j)},\bm{\varpi}^{(j)})$ for $p=1,2,3$. 
\EndIf
\EndIf
\State
Update the iteration number $j\leftarrow j+1$.
\EndWhile
\State
$(\lambda^*,\mu^*,\alpha^*,\mathbf{D}^*) = (\lambda^{(j)},\mu^{(j)},\alpha^{(j)},\Db^{(j)})$.
\end{algorithmic}
\end{algorithm}

In this section we describe the algorithmic details of our learning approach and describe the learning workflow that, starting with MD displacements, delivers the optimal influence function $K$ and the material parameters.

Numerically, the constrained optimization problem \eqref{eq:true_problem_full} poses several challenges. Firstly, the quadrature weights $W_{j,i}$ generated in the preprocessing generally depend on the horizon size $\delta$, which hinders the application of a suite of continuous optimization techniques such as gradient descent or Adam. A similar issue applies to $M$. Moreover, due to the solvability constraints, \eqref{eq:true_problem_full} is expected to be nonconvex and likely to exhibit local minima. Lastly, the numerical evaluations of the eigenvalues are time-consuming. For all these reasons, for the sake of numerical efficiency, we treat $\delta$ and $M$ as hyperparameters to be separately tuned to achieve the best learning accuracy without overfitting. As suggested by \cite{You2020Regression},  the optimization problem \eqref{eq:true_problem_full} is split into a {\it prediction} step (without constraints) and a {\it correction} step (with constraints), and propose a ``two-stage'' strategy, whose key components are summarized in Algorithm \ref{alg:augmented_lagrangian}.

The prediction step of the algorithm relies on the fact that, as shown in \cite{mengesha14Navier}, when $\alpha<3$, $\lambda+\mu>0$, $\mu>0$ and $K\geq0$, the LPS problem is guaranteed to be well-posed, and therefore no additional solvability constraint of $K$ are required. Therefore, in the prediction step, we find a set of nonnegative Bernstein coefficients that will be used as an initial guess for the second, correction step. The nonnegative coefficients and the corresponding nonnegative influence function are denoted by $\Db^{pre}_k$ and $K^{pre}$, respectively. These are obtained by solving the following, unconstrained problem, whose full solution is denoted by $(\lambda^{pre},\mu^{pre},\alpha^{pre},\Db^{pre})$.
\begin{align*}
&L^{pre}(\lambda,\mu,\alpha,\Db)
=\frac{1}{S} \sum_{s=1}^S \sum_{\xb_i\in\mcX^s}\left| \bb^s_i+ \sum_{k=0}^{M} \frac{D_k}{m_i(\delta)}\sum_{ \mathbf{x}_j \in B_\delta (\mathbf{x_i})} \frac{B_{k,M}\left (\frac{|\mathbf{x}_j-\mathbf{x}_i|}{\delta} \right)W_{j,i}}{|\mathbf{x}_j-\mathbf{x}_i|^{\alpha}} \right.\\
    &\left.\left[C_1\left(\lambda- \mu\right)  \left(\mathbf{x}_j-\mathbf{x}_i \right)\left(\theta^s_i + \theta^s_j \right) +C_2 \mu \frac{\left(\mathbf{x}_j-\mathbf{x}_i\right)\otimes\left(\mathbf{x}_j-\mathbf{x}_i\right)}{\left|\mathbf{x}_j-\mathbf{x}_i \right|^2}  \left(\mathbf{u}^s_j - \mathbf{u}^s_i \right)\right] \right|^2,
\end{align*}
where
\begin{align*}
&\theta^s_i:=\sum_{k=0}^{M} \frac{2D_k}{m_i(\delta)} \sum_{ \mathbf{x}_j \in B_\delta (\mathbf{x}_i)}  \frac{B_{k,M}\left (\frac{|\mathbf{x}_j-\mathbf{x}_i|}{\delta} \right)W_{j,i}}{|\mathbf{x}_j-\mathbf{x}_i|^{\alpha}} (\mathbf{x}_j-\mathbf{x}_i)\cdot \left(\mathbf{u}^s_j - \mathbf{u}^s_i \right),\\
&m_i(\delta):=\sum_{k=0}^{M} {D_k}\underset{\mathbf{x}_j \in B_\delta (\mathbf{x}_i)}{\sum} \frac{B_{k,M}\left (\frac{|\mathbf{x}_j-\mathbf{x}_i|}{\delta} \right)W_{j,i}}{|\mathbf{x}_j-\mathbf{x}_i|^{\alpha}}\verti{\mathbf{x}_j-\mathbf{x}_i}^2.
\end{align*}
We solve this unconstrained optimization problem via the Adam optimizer \cite{kingma2014adam} and we guarantee that $\Db\geq0$ by using the map 
$
\mathbf{D} \mapsto \text{ReLU}(\mathbf{D})
$
after each step of gradient descent. Here \text{ReLU} denotes the rectified linear unit function:
\begin{displaymath}
\text{ReLU}(x):=\left\{\begin{array}{cc}
     0,& \text{ for }x\leq 0; \\
     x,& \text{ for }x>0.
\end{array}\right.
\end{displaymath}
This operation does not modify the value of $L^{pre}$ and ensures that the sequence of iterates and the local minimum are nonnegative. Similar strategies are also applied to $\lambda$, $\mu$ and $\alpha$ as follows\footnote{Note that although we require $\lambda+\mu>0$, $\mu>0$ and $\alpha<3$ for the well-posedness analysis in Theorem \ref{lem:suf}, for numerical simplicity we employ strategy \eqref{eqn:relualpha} which only guarantees $\lambda+\mu\geq0$, $\mu\geq0$ and $\alpha\leq3$. However, in all numerical tests we observed that the predicted $\lambda$, $\mu$, $\alpha$ satisfies the well-posedness conditions, as will be shown in Section \ref{sec:MD}.}:
\begin{equation}\label{eqn:relualpha}
\begin{aligned}
&\mu \mapsto \text{ReLU}(\mu),\\
&\lambda \mapsto \text{ReLU}(\lambda+\mu)-\mu,\\
&\alpha\mapsto 3-\text{ReLU}(3-\alpha).
\end{aligned}
\end{equation}

The second step of the algorithm corrects the prediction-step solution by solving for the fully constrained optimization problem. Specifically, by employing $(\lambda^{pre},\mu^{pre},\alpha^{pre},\Db^{pre})$ as the initial guess, we apply the augmented Lagrangian method \cite{yu2019dag,nocedal2006numerical} and treat the inequality constraints via slack variables. To do this, introduce the functions
\begin{equation}\label{eq:ineq} 
\left\{
\begin{aligned}
\quad H_1(\alpha,\mathbf{D},\varpi_1)= &\Lambda(\Gamma_{(\alpha,\Db,\delta,M)})-\zeta-\varpi_1^2,\\
\quad H_2(\alpha,\mathbf{D},\varpi_2)=&\Lambda(\Phi_{(\alpha,\Db,\delta,M)}\Gamma^{+}_{(\alpha,\Db,\delta,M)}\Phi^t_{(\alpha,\Db,\delta,M)})- \zeta-\varpi_2^2, \\
\quad H_3(\alpha,\mathbf{D},\varpi_3)=&\Lambda(\Gamma_{(\alpha,\Db,\delta,M)}-2\Phi^t_{(\alpha,\Db,\delta,M)}\Phi_{(\alpha,\Db,\delta,M)})-\varpi_3^2.
\end{aligned}
\right.
\end{equation}
Here, $\bm{\varpi}=[\varpi_1,\varpi_2,\varpi_3]$ is the vector of slack variables arising from the inequality constraints. Then, minimize the following penalized loss function:
\begin{align}\label{eq:L2}
&L^{corr}(\lambda,\mu,\alpha,\mathbf{D}, \bm{\varpi})
=\frac{1}{S} \sum_{s=1}^S \sum_{\xb_i\in\mcX^s}\left| \bb^s_i+ \sum_{k=0}^{M} \frac{D_k}{m_i(\delta)}\sum_{ \mathbf{x}_j \in B_\delta (\mathbf{x_i})} \frac{B_{k,M}\left (\frac{|\mathbf{x}_j-\mathbf{x}_i|}{\delta} \right)W_{j,i}}{|\mathbf{x}_j-\mathbf{x}_i|^{\alpha}} \right.\\
    &\left.\left[C_1\left(\lambda- \mu\right)  \left(\mathbf{x}_j-\mathbf{x}_i \right)\left(\theta^s_i + \theta^s_j \right) +C_2 \mu \frac{\left(\mathbf{x}_j-\mathbf{x}_i\right)\otimes\left(\mathbf{x}_j-\mathbf{x}_i\right)}{\left|\mathbf{x}_j-\mathbf{x}_i \right|^2}  \left(\mathbf{u}^s_j - \mathbf{u}^s_i \right)\right] \right|^2\\
&+\sum_{p=1}^3\phi_p H_p(\alpha,\mathbf{D},\bm{\varpi})+\dfrac{\psi}{2}\sum_{p=1}^3H_p^2(\alpha,\mathbf{D},\bm{\varpi}),
\end{align}
where $\phi_p$, $p=1,2,3$ are the Lagrange multipliers and $\psi$ is a penalty parameter. At this stage, the Adam optimizer is used. 
An iterative procedure updates the Lagrange multipliers by 1) solving the unconstrained optimization problem, 2) updating $\phi_p$ via dual gradient ascent
$$\phi_p\mapsto \phi_p+\psi H_p(\alpha,\Db,\bm{\varpi}),$$
and 3) increasing $\psi$ when the decreasing ratio of $H_p(\alpha,\Db,\bm{\varpi})$ (with respect to the last iteration)  does not reach $4$ for at least one $p\in\{1,2,3\}$. 
As done in the prediction step, at each iteration of the gradient descent, we map $\mu$ to ReLU$(\mu)$, $\lambda$ to ReLU$(\lambda+\mu)-\mu$, and $\alpha$ to $3-\text{ReLU}(3-\alpha)$. Further details on parameter updates are described in Algorithm \ref{alg:augmented_lagrangian}. The optimal solution is denoted by $(\lambda^*,\mu^*,\alpha^*,\mathbf{D}^*)$.


In applying Algorithm \ref{alg:augmented_lagrangian}, in all our tests, we run the Adam optimizer in PyTorch using a batch size of $70$, the inequality constraint tolerance is set to $\epsilon$=1E-5 and the learning rate to 1E-3. The algorithm runs until the loss stagnates, indicating that a stationary point has been reached. Stagnation typically happens between 1000 and 2000 epochs for the first stage of the algorithm and at $\sim$500 epochs for the second stage at each iteration of the augmented Lagrangian method. 
 

\begin{algorithm}
\caption{Workflow for learning the operator $\mathcal{L}_K$ from MD displacements.}\label{alg:workflow}
\begin{algorithmic}[1]
\State Generate MD displacements on fine grids $\{X_\veps^s\}$ using different external forcings and domains configurations and group the samples in three data sets:
$$\mathbb{MD}_{{train/val/test}}:=\{M^s_{\veps,s},\Ub^s_{\veps,s}(t),\Bb^s_{\veps,s}(t)\},\,s=1,\cdots,S_{train/val/test}.$$
\State
Smooth the data sets $\mathbb{MD}_{train/val/test}$ in space and time and evaluate the smoothed data at coarser grids $\mcX^s$ to obtain the sets
$$\mathbb{S}_{train/val/test}:=\{\ub^s(\xb_{i,s}),\bb^s(\xb_{i,s})\},\,s=1,\cdots,S_{train/val/test}.$$
\For{$M\in \mathbb{M}$:} 
\For{$\delta\in \mathbb{D}$:}
\State
Perform the two-stage optimization strategy for fixed $(\delta,M)$ to obtain $$(\lambda^*_{(\delta,M)},\mu^*_{(\delta,M)},\alpha^*_{(\delta,M)},\mathbf{D}^*_{(\delta,M)}).$$
\EndFor
\State
Find $\delta^*_M$ that minimizes ${\rm Res}(\delta,M;\mathbb S_{train})$.
\State
Calculate and store $E^{train}_{\rm Res}(\delta^*_M,M)$, $E_\ub^{train}(M)$, $E^{val}_{\rm Res}(M)$ and $E_\ub^{val}(M)$.
\EndFor
\State
Find $M^*$ that minimizes the average of the normalized errors in step 7 and set
$$\begin{aligned}
(\lambda^*,\mu^*,\alpha^*,\mathbf{D}^*,\delta^*,M^*) 
&= (\lambda^*_{(\delta^*_{M^*},M^*)},\mu^*_{(\delta^*_{M^*},M^*)},\alpha^*_{(\delta^*_{M^*},M^*)},\mathbf{D}^*_{(\delta^*_{M^*},M^*)},\delta^*_{M^*},M^*)\\
\mcL^*_{K}&=\mcL_{K(\delta^*_{M^*},M^*)}
\end{aligned}$$

\end{algorithmic}
\end{algorithm}

We now discuss how to tune $\delta$ and $M$, which are treated as hyperparameters. Divide the sample set into two sub-sets: the set of training samples $\mathbb{S}_{train}:=\{\ub^s(\xb_{i,s}),\bb^s(\xb_{i,s})\}$, $s=1,\cdots,S_{train}$, and the set of validation samples $\mathbb{S}_{val}:=\{\widetilde{\ub}^s(\xb_{i,s}),\widetilde{\bb}^s(\xb_{i,s})\}$, $s=1,\cdots,S_{val}$. While all samples correspond to the same material, the problem setting of training and validation samples can be substantially different. For example, they could be generated with different grids $\mcX^s$, loading scenarios, boundary conditions, and geometric configuration. For $M\in\mathbb M$, we perform Algorithm \ref{alg:augmented_lagrangian} using the training set $\mathbb{S}_{train}$ with different values of $\delta\in\mathbb D$ and denote the corresponding nonlocal operator by $\mathcal{L}_{K(\delta,M)}$. Then, for each $M$, define the optimal horizon $\delta^*_M$ as the one that minimizes the average residual of the LPS equation, denoted by $E_{\rm Res}^{train}$. Formally,
$$
\delta^*_M=\underset{\delta\in\mathbb D}{\text{argmin }}\; E_{\rm Res}^{train}(\delta,M)=\underset{\delta}{\text{argmin }}\; {\rm Res}(\delta,M;\mathbb S_{train}),
$$
where
\begin{align}\label{eq:L_train}
{\rm Res}(\delta,M;\mathbb S_{train}):=
\frac{1}{S_{train}} \sum_{\{\ub^s(\xb_{i,s}),\bb^s(\xb_{i,s})\}\in\mathbb{S}_{train}} ||\mcL_{K(\delta,M)}^h\ub^s(\xb_{i,s}) - \bb^s(\xb_{i,s}) ||^2_{\ell_2(\mcX^s)}.
\end{align}
Denote the corresponding nonlocal operator by $\mathcal{L}_{K(\delta^*_M,M)}$. 
Next, to determine the optimal $M$ that allows for accurate representation of the training samples without overfitting, we test, for each $M$, the optimal operator $\mathcal{L}_{K(\delta^*_M,M)}$ on the validation set $\mathbb S_{val}$ by evaluating the average residual of the LPS equation; the latter, denoted by $E_{\rm Res}^{val}$, is defined as
\begin{align}\label{eq:L_val}
E_{\rm Res}^{val}(M)={\rm Res}(\delta^*_M,M;\mathbb S_{val}).
\end{align}
In principle, small training and validation losses indicate that the model performs well on the training set and generalizes well to other data sets. {As an additional metric of accuracy on both the training and validation sets, also consider the displacement mean square error (MSE), denoted by $E^{train}_\ub$ and $E^{val}_\ub$, respectively. Formally,
\begin{align}\label{eq:error_val}
E_\ub^{train}(M):=
\frac{1}{S_{train}} \sum_{\{\ub^s(\xb_{i,s}),\bb^s(\xb_{i,s})\}\in\mathbb{S}_{train}} ||\ub^s(\xb_{i,s}) - (\mcL_{K(\delta^*_{M},M)}^h)^{-1}\bb^s(\xb_{i,s}) ||^2_{\ell_2(\mcX^s)},
\end{align}
and $E^{val}_\ub$ is defined similarly by taking the average over the validation set $\mathbb{S}_{val}$.} 
Based on these metrics, we select $M$ such that the average of the normalized $E^{train}_{\rm Res}(\delta^*_M,M)$, $E_\ub^{train}(M)$, $E^{val}_{\rm Res}(M)$ and $E_\ub^{val}(M)$ is minimized. Here the normalization is taken with respect to the same quantities evaluated at the baseline, which is the case with $M=0$ (the constant Bernstein polynomial). In particular, take $M^*=\underset{M}{\text{argmin}}\,\text{AvgE}(M)$ where
$$\text{AvgE}(M):=\frac{1}{4} \left(\frac{E^{train}_{\rm Res}(\delta^*_M,M)}{E^{train}_{\rm Res}(\delta^*_0,0)},\frac{E^{train}_{\ub}(\delta^*_M,M)}{E^{train}_{\ub}(\delta^*_0,0)},\frac{E^{val}_{\rm Res}(\delta^*_M,M)}{E^{val}_{\rm Res}(\delta^*_0,0)},\frac{E^{val}_{\ub}(\delta^*_M,M)}{E^{val}_{\ub}(\delta^*_0,0)}\right).$$

A summary of the above strategy can be found in Algorithm \ref{alg:workflow} where we report the overall workflow of our learning procedure. 


\section{Consistency tests for manufactured solutions}\label{sec:consistency}

\begin{figure}
    \centering
    \includegraphics[width = .48\textwidth]{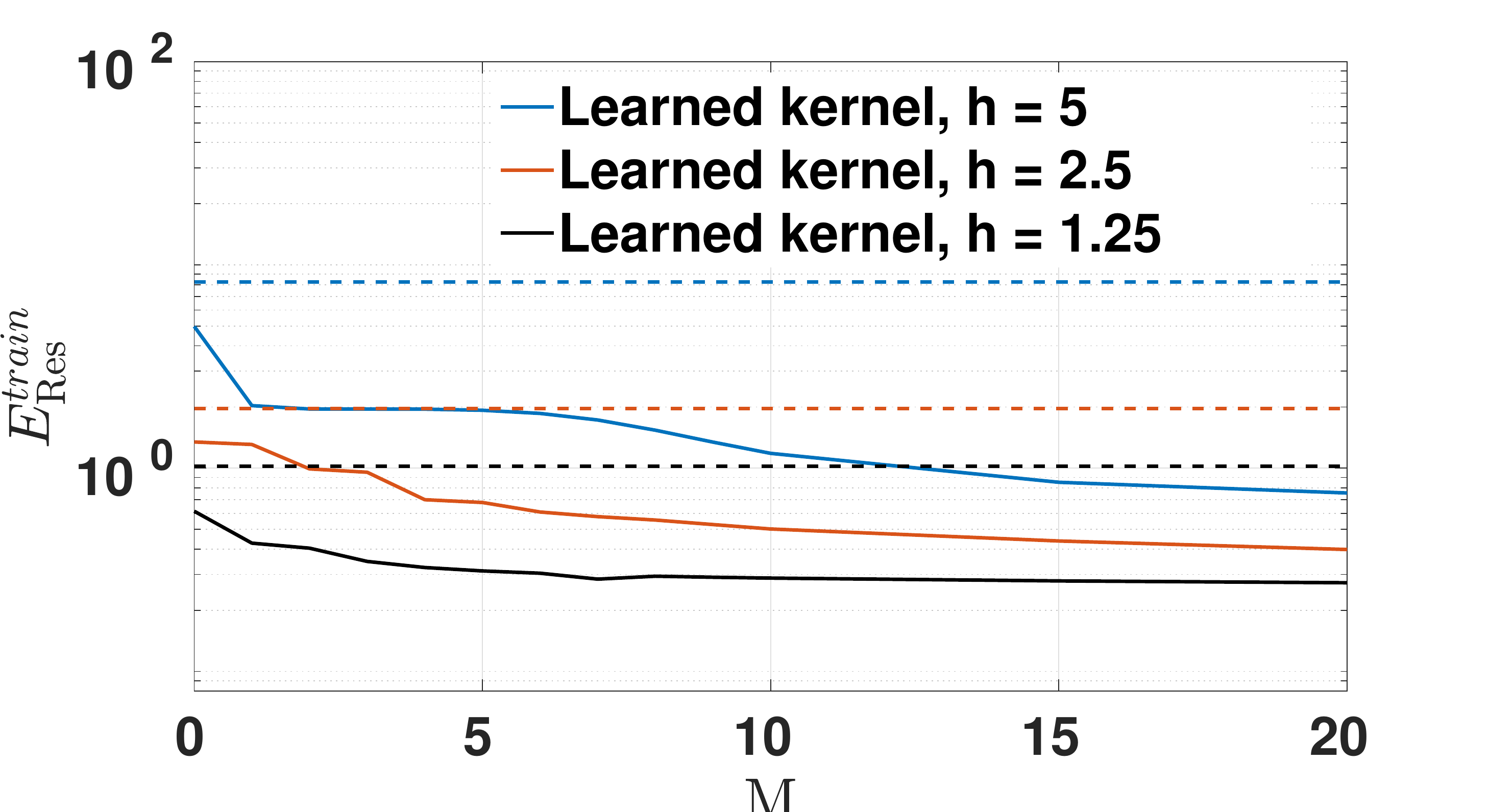}
    \includegraphics[width = .48\textwidth]{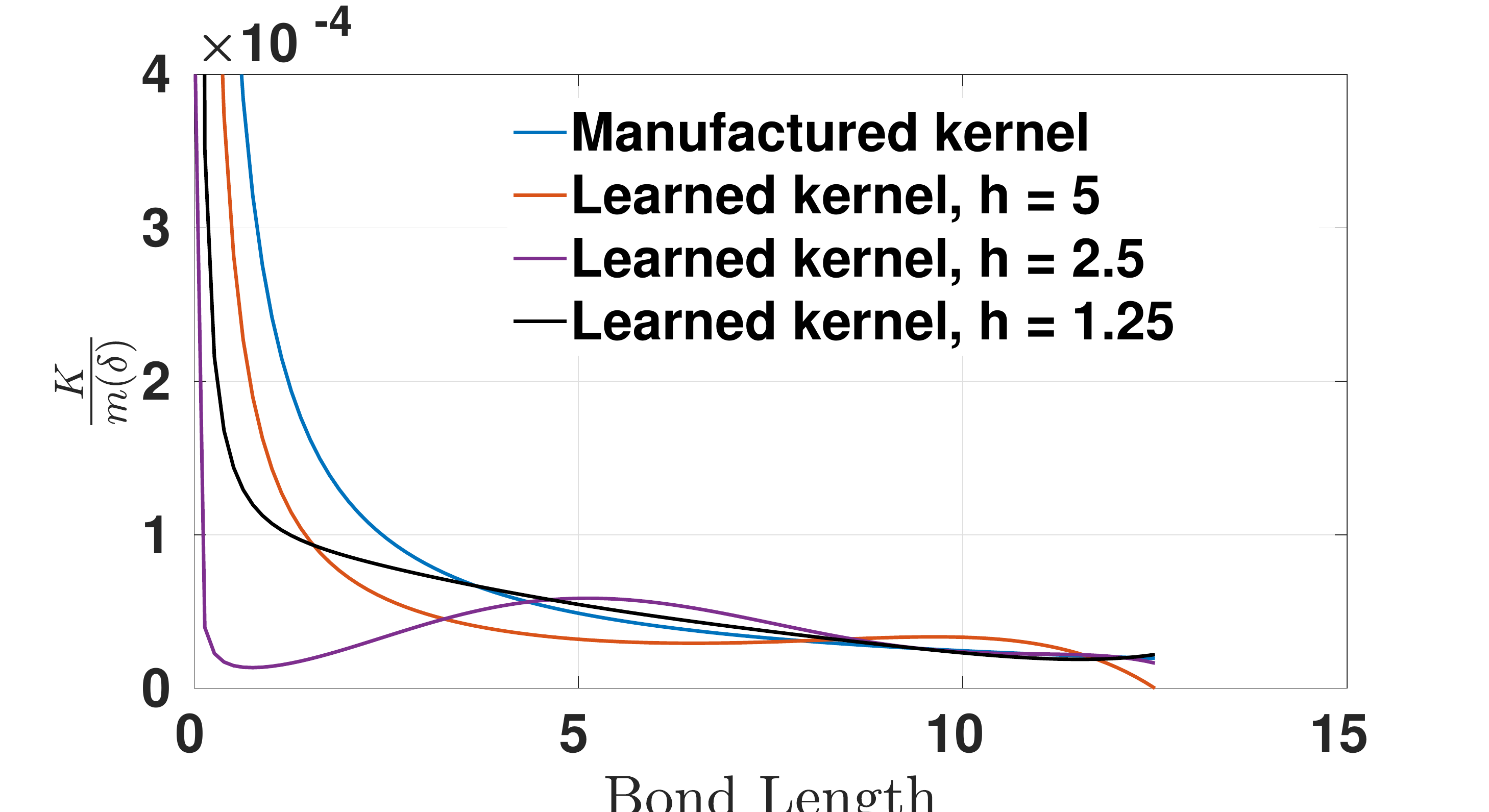}
    \caption{Consistency tests for Algorithm \ref{alg:augmented_lagrangian} with fixed $\alpha=1$ and positive influence function. Left: The training loss versus basis order M when using different grid size; the dashed lines indicate the values of the loss functions when the manufactured kernel is used, colors reflect the values of $h$ used for discretization. Right: The comparison of learned influence functions and the manufactured influence function $K_{\rm man} = \frac{1}{r}$.}
    \label{fig:reproduce_const}
\end{figure}

\begin{figure}
    \centering
    \includegraphics[width = .48\textwidth]{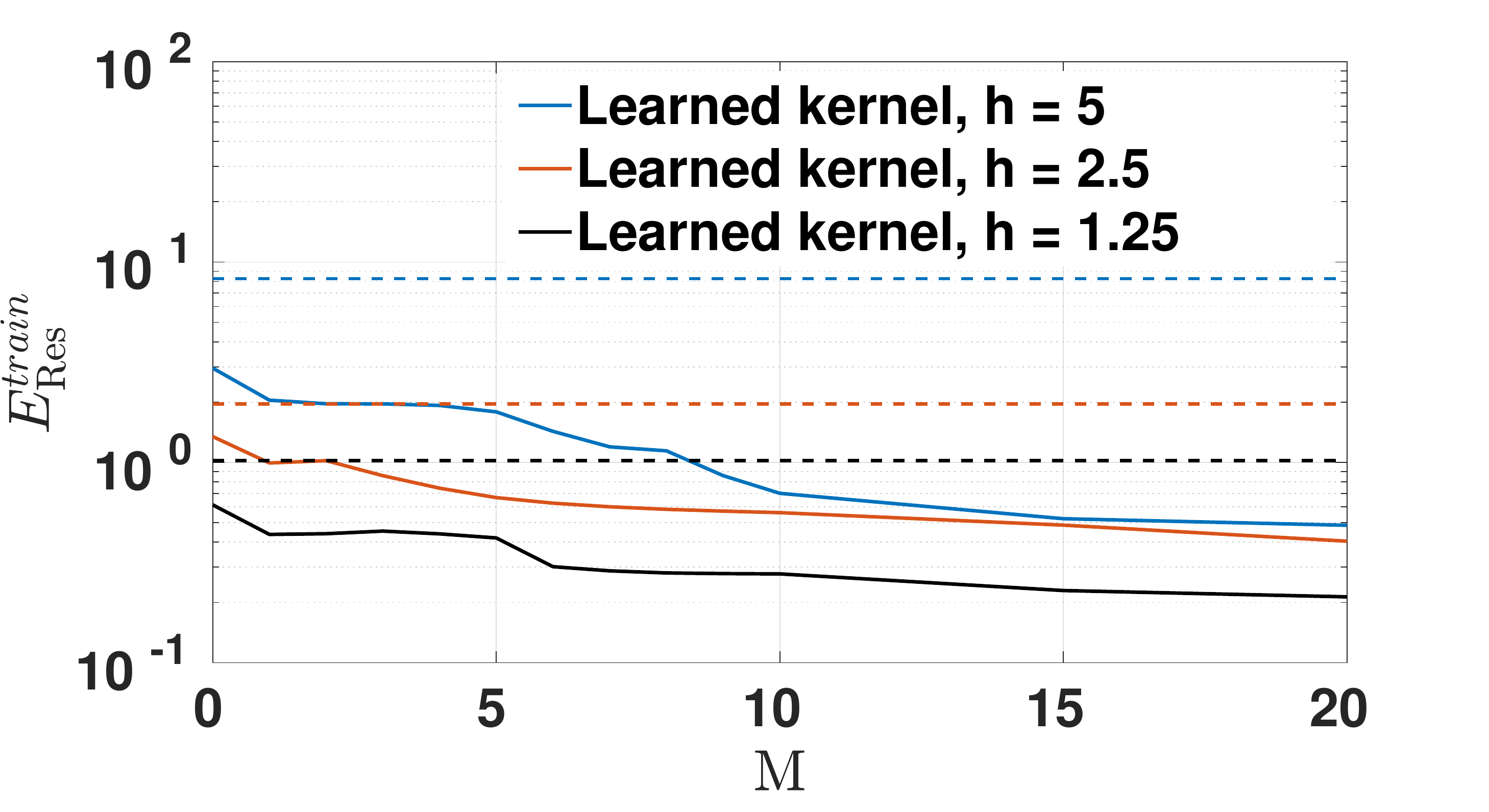}
    \includegraphics[width = .48\textwidth]{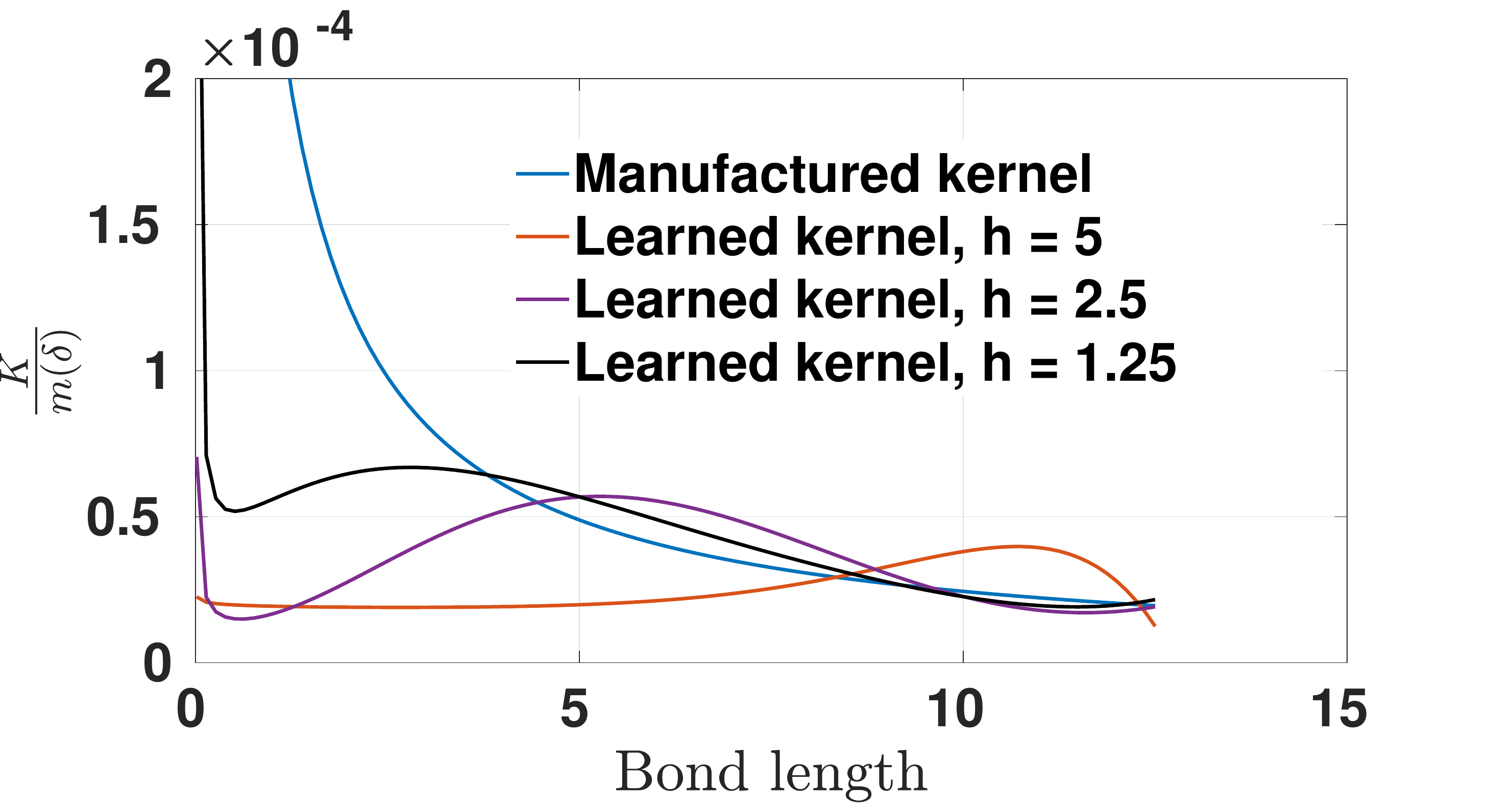}
    \caption{Consistency tests for Algorithm \ref{alg:augmented_lagrangian} and positive influence function, while learning $\alpha$. Left: The training loss versus basis order M when using different grid size; the dashed lines indicate the values of the loss functions when the manufactured kernel is used, colors reflect the values of $h$ used for discretization. Right: The comparison of learned influence functions and the manufactured influence function $K_{\rm man} = \frac{1}{r}$.}
    \label{fig:reproduce_alpha}
\end{figure}

In this section, we test our operator regression algorithm by considering analytic training pairs $\{\ub^s(\xb),\bb^s(\xb)\}$ satisfying \eqref{eq:nonlocElasticity} for the manufactured influence function
\begin{equation}\label{eq:kernel_man}
    K_{\rm man}(\xb,\yb) = \frac{1}{\verti{\xb-\yb}}.
\end{equation}
In particular, data sets are considered with different spatial resolutions $\mcX_h:=\{(p_1h,p_2h)|\bm{p}=(p_1,p_2)\in\mathbb{Z}^2\}\cap(\Omega\cap\Omega_I)$ and different Bernstein polynomial orders $M$ with the purpose of validating Algorithm \ref{alg:augmented_lagrangian} before employing it on MD data sets.

For the computational domain $\Omega=[0,100]^2$, the training data pairs $\{ \ub^s(\xb), 
{\bf b}^s(\xb)\}_{s=1}^{70}$ are generated by setting
\begin{equation}
\begin{aligned}
\ub(x_1,x_2) & = (0.1\cos(2k_1 \pi x_1)\cos( 2k_2 \pi x_2), 0), \; \text{or} \\
\ub(x_1,x_2) &= (0, 0.1\cos(2k_1 \pi x_1) \cos(2k_2 \pi x_2)),
\end{aligned}
\end{equation}
with $k_1,k_2 \in \{0,1,2,3,4,5\}$. Then, for each displacement field $\ub^s(\xb)$, the corresponding forcing field ${
\bf b}^s$ is computed from \eqref{eq:nonlocElasticity} with $\lambda = 0.1010$, $\mu =0.4545$, and $\delta = 0.125$. By evaluating $\ub^s$ and $\bb^s$ on different grid sets $\mathcal{X}_h$ with $h=5$, $h=2.5$ and $h=1.25$, respectively, three training sets of size 70 are then obtained. These are denoted by $\mathbb{S}^{h=5}_{train}$, $\mathbb{S}^{h=2.5}_{train}$ and $\mathbb{S}^{h=1.25}_{train}$.

To verify the consistency of the prediction step in Algorithm \ref{alg:augmented_lagrangian}  and its behavior with respect to increasing resolution and polynomial order, first consider a positive influence function $K$ with prescribed fractional order $\alpha =1$. Then choose the Bernstein basis order $M$ in $[0,20]$. The prediction step involves the solution of a convex optimization problem with a non-empty feasible set. Therefore, every local minimum is a global minimum. 

For different training sets, the training losses $E_{\rm Res}^{val}(M)$ are plotted with respect to increasing polynomial orders $M$ in Figure \ref{fig:reproduce_const}, left. These results suggest that the training loss improves as the basis order $M$ is increased and as the grid size $h$ decreases. Furthermore, the manufactured ground-truth kernels have a higher training losses compared to the learned kernels due to the discretization error, which suggests that the learning algorithm is able to obtain better kernels on each grid set. Figure \ref{fig:reproduce_const}, right, shows a comparison of the learned influence functions for a fixed polynomial order $M=10$. The learned influence function gets closer to the manufactured influence function $K_{man}$ as $h\rightarrow 0$. This illustrates the consistency of the learning algorithm for a given $\alpha$.

To investigate the effects of the fractional order $\alpha$, we now consider a positive influence function with unknown fractional order and use, again, the prediction step of Algorithm \ref{alg:augmented_lagrangian}. In this case, the convergence of the optimal kernel to the manufactured one is not guaranteed, since the fractional order is a tunable parameter. 
The training losses and learnt influence functions are provided in Figure \ref{fig:reproduce_alpha}. Although the algorithm does not recover exactly the influence function $K_{man}$, low values of $E^{train}_{\rm Res}$ can be achieved as $M$ increases.

\section{Application to graphene using MD}\label{sec:MD}
\begin{figure}  
\centering
\includegraphics[width=0.8\textwidth]{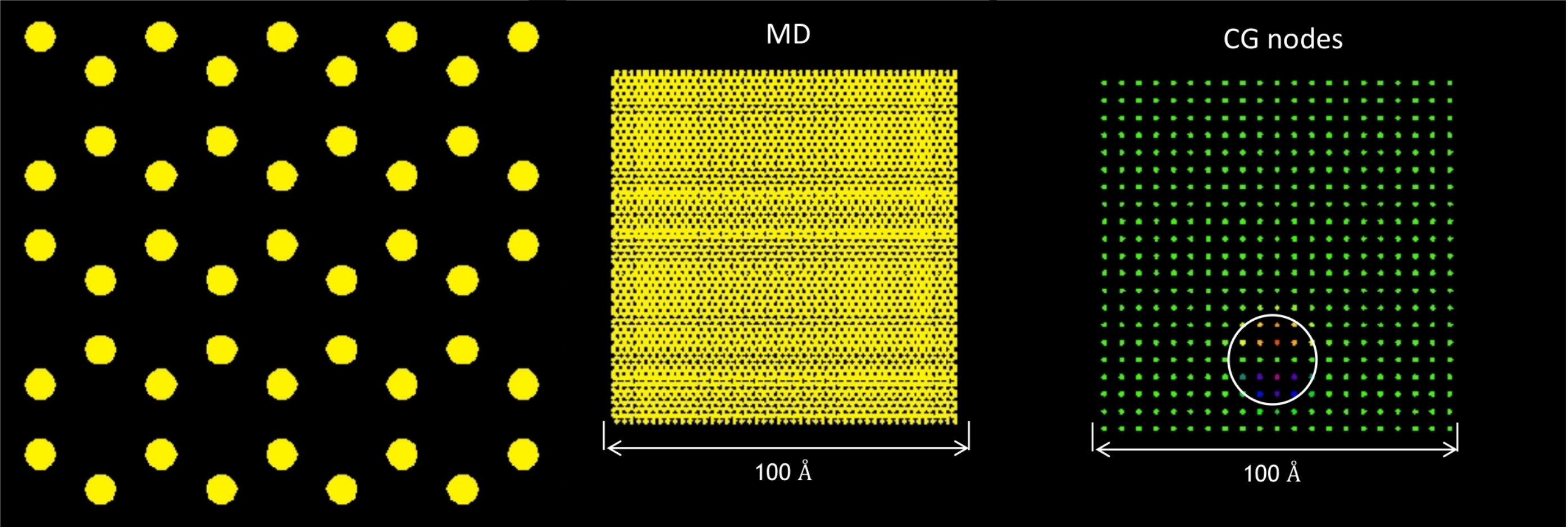}
\caption{Left: hexagonal graphene atomic structure.
Center: full MD grid.
Right: coarse grained node positions.}
\label{fig-atoms}
\end{figure}

To illustrate the efficacy of our method in obtaining an optimal peridynamic model from coarse-grained MD displacements, we consider graphene sheets as the application.
Graphene is a two-dimensional form of carbon with a hexagonal structure.
Because of its high stiffness and strength, as well as other unusual physical properties, graphene is being studied for possible use in a number of applications, including as a structural material.
For the present study, an MD model was created using the Tersoff interatomic potential \cite{tersoff88}.
This potential is widely used in the MD community for graphene because it incorporates the relative rotation angle between covalent bonds, which strongly affects the mechanical response.
A thermostat is included in the MD model in the present study to control the temperature and periodic boundary conditions are applied.
The MD grid is shown in Figure~\ref{fig-atoms}, center. The grid has 3588 atoms. The corresponding coarse-grained node positions are shown in Figure~\ref{fig-atoms}, right.
To simplify the analysis, out-of-plane motions were not considered in this study, although they would occur in a real material.

Unstressed graphene nominally has an interatomic spacing of 1.46\AA.
In this study, values of the coarse-grained quantities $\ub_i$ and $\bb_i$ are evaluated on a square lattice of nodes indexed by
$i$ with spacing $h$=5.0\AA. {We also consider an additional, finer data set generated for validation purposes with spacing 2.5\AA}.

{For any coarse grained node position $\xb_i$
and any atomic position $\Xb_\veps$, define the smoothing function by
\begin{equation}
    \omega(\xb_i,\Xb_\veps) = \frac{\tau(\xb_i,\Xb_\veps)}{\sum_j \tau(\xb_j,\Xb_\veps)}
    \label{eqn-omegaMD}
\end{equation}
where the cone-shaped function
$\tau(\xb,\Xb) = \max\big\{0,\;R-|\Xb-\xb|\big\}
$
induces a coarse-graining radius of $R$=10.0\AA. Note that \eqref{eqn-omegaMD} satisfies the normalization requirement \eqref{eqn-omnorm}.}

In all cases, external loading is applied to the atoms in the MD grid in addition to the random loads applied by the thermostat.
The external loading for each atom $\Bb_\veps$ is constant over time. The magnitude of the loading is chosen so that the bond strains are no larger than 2\%, 
which is less than the strains at which nonlinear effects appear.
As described in Section \ref{sec:noise}, the magnitude of the loading is varied relative to the forces on the atoms that sustain the thermal oscillations. This variation helps to test the robustness of the machine learning method in extracting the continuum material properties in the presence of {thermal oscillations that create noise.}

\subsection{Data sets and metrics of accuracy}

Three sets of data are generated from the MD simulations for each of two values of temperature, $0K$ and $300K$.
In all MD experiments, the atoms are initialized with positions on a hexagonal lattice in the $x_1$-$x_2$ plane with an interatomic spacing of $1.46\AA$.
The mass of each atom is 2.0E-26kg, or 12amu.
For purposes of computing stresses, the thickness of the lattice is set to 3.35\AA, which is the approximate distance between layers in multilayer graphene.
The MD time step size is 5.0E-16s, or 5.0fs.

\begin{figure}
    \centering
    \includegraphics[width=0.8\textwidth]{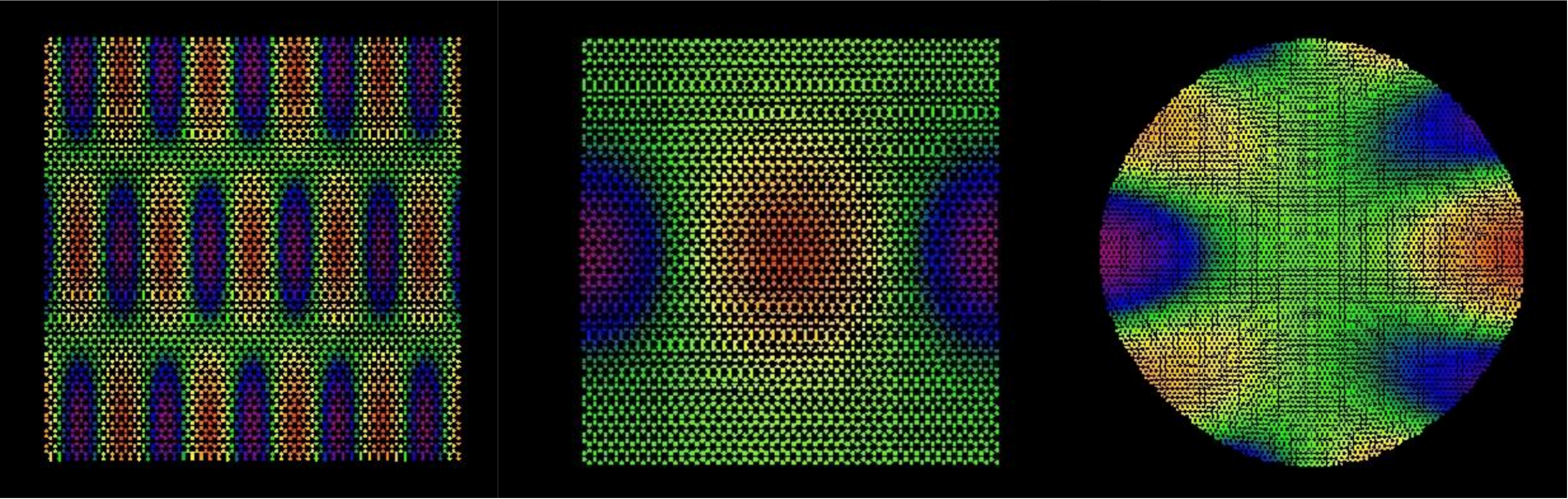}
    \caption{Contours of $U_1$ displacement in typical MD simulations at zero temperature for the three types of datasets. 
    Left: training. Center: validation. Right: testing.}
    \label{fig-mddispl}
\end{figure}

Two types of samples are generated for each data set: the standard samples with spacing $h$=5.0\AA\, which will be denoted by $\mathbb{S}^{\text{0K/300K}}_{train/val/test}$, and the samples with finer grids $h$=2.5\AA\,, denoted by $\mathbb{S}^{\text{0K/300K,fine}}_{train/val/test}$. Unless stated otherwise, the $h$=5.0\AA\, data sets are used in the learning tasks. The finer grid data sets are employed to assess the generalization properties of the proposed learning approach to different grids (further details and discussions are provided in Section \ref{sec:generalization}).
Images showing contours of $U_1$, the component of atomic displacement in the $x_1$ direction, for the training, validation, and testing samples are shown in Figure~\ref{fig-mddispl}.

\medskip\noindent
{\it 1) Training data set (70 samples):} The MD domain is a $100$\AA $\times 100$\AA$\;$ square, and, for {$k_1,k_2\in\{0, \frac{\pi}{50},\frac{2\pi}{50},\ldots,\frac{5\pi}{50}\},$} the prescribed external loadings are given by
\begin{equation}\label{eqn:s_train_300K}
\bb(x_1,x_2) = (C^1_{k_1,k_2}\cos(k_1x_1)\cos(k_2x_2),0), \text{ or } \bb(x_1,x_2) = (0,C^2_{k_1,k_2}\cos(k_1x_1)\cos(k_2x_2)).
\end{equation}
As mentioned above, the constant $C^1_{k_1,k_2}$ and $C^2_{k_1,k_2}$ are adjusted so that the bond strains are no larger than 2\%, so the deformation remains in the linear range of material response.

\medskip\noindent
{\it 2) Validation data set (10 samples).} 
For the same MD grid and coarse-grained nodes as in the training data set, the applied loads in the validation data set are as follows:
\begin{equation}\label{eqn:val-load}
\bb(x_1,x_2) = (C_k^1,C_k^2) \sum_{j=-1}^1 (-1)^j 
  \cos\left( \frac{\pi}{2}\min\left\{1,\frac{r_{j,k}}{R_k}\right\} \right)
\end{equation}
where
\begin{equation}\label{eqn:val-rj}
  r_{j,k}=\sqrt{ (x_1-(1-p_k)Lj)^2+(x_2-p_kLj)^2 }
\end{equation}
where $L$=50
and the values of the parameters $C_k^1$, $C_k^2$, $p_k$ and $R_k$ are given in Table~\ref{table-valparam}.
In each case, loads are applied to the atoms within three disks of radius $R_k$ with centers at the
center of the grid and at the left and right boundaries (if $p_k=0$) or the upper and lower boundaries (if $p_k=1$).
The direction of the load vectors is either in the $x_1$ or $x_2$ directions.
The loads in all cases are self-equilibrated and periodic.

\begin{table}  [t]
\centering
    \begin{tabular}{|c|c|c|c|c|} \hline
        $k$ & $C_k^1$     & $C_k^2$  & $p_k$ & $R_k$ \\  \hline\hline
        1   & 0.001       & 0        & 0     & 25    \\  \hline
        2   & 0           & 0.001    & 0     & 25    \\  \hline
        3   & 0.001       & 0        & 0     & 15    \\  \hline
        4   & 0           & 0.001    & 0     & 15    \\  \hline
        5   & 0.001       & 0        & 0     & 10    \\  \hline
        6   & 0.001       & 0        & 1     & 25    \\  \hline
        7   & 0           & 0.001    & 1     & 25    \\  \hline
        8   & 0.001       & 0        & 1     & 15    \\  \hline
        9   & 0           & 0.001    & 1     & 15    \\  \hline
        10  & 0.001       & 0        & 1     & 10    \\  \hline
    \end{tabular}
    \caption{Parameters used in the MD loading in the 10 validation tests.}
  \label{table-valparam}
\end{table}

\medskip\noindent
{\it 3) Test data set (4 samples).}
To demonstrate that the learned material model applies to geometries different from the original square geometry, four additional test cases are considered. Here, the MD region is a disk of radius 100\AA.
Within this disk loading is applied as listed in Table~\ref{table-testload} to the exterior
of a circle with radius 50\AA, with the interior unloaded.
The equilibrium displacements, with smoothing as described previously, are computed at
the coarse-grained nodes, which are spaced 5\AA\, or 2.5\AA, apart on a square lattice.
All of these cases have loadings that are discontinuous functions of the radius and 
therefore are more challenging from a modeling perspective than the validation cases described above.
In cases 2 and 4 the loading is also a discontinuous function of the angle because of the sign function (sgn).

\begin{table}  [t]
\centering
    \begin{tabular}{|c|l|l|} \hline
        Case  & $b_1$ & $b_2$\\  
        \hline\hline
        1     & $C\cos4 \theta \cos \theta  $               & $ C\cos4\theta \sin\theta $ \\  \hline
        2     & $C\,{\mathrm{sgn}}(\cos4\theta) \cos \theta $ &  $C\, {\mathrm{sgn}}(\cos4\theta)\sin\theta $ \\  \hline
        3     & 0                                           &  $C\, {\mathrm{sgn}}(\sin\theta)\sin\theta $  \\  \hline
        4     & $C\sin3\theta \sin \theta       $         &  $C \sin3\theta\cos\theta$  \\  \hline
    \end{tabular}
    \caption{Loading applied to the exterior of a disk of radius 50\AA\, in the four tests. In all cases, $C$=0.0005. $b_1$ and $b_2$ are components of $\bb$ along the $x_1$ and $x_2$ directions, respectively, and 
    $\theta$ is the polar angle in the plane.}
  \label{table-testload}
\end{table}

As metrics of accuracy on the training, validation and test sets, in this section we calculate the averaged mean square loss (MSL) and displacement mean square error (MSE) for each learnt kernel on these three sets, which will be referred to as $E^{train/val/test}_{Res/\ub}$, respectively. To provide a fair comparison between different sets, all these accuracy metrics except $E^{test}_{Res}$ are normalized with respect to either the force loading or the displacement fields. Specifically, $E^{train/val}_{Res}$ is normalized by $\vertii{\bb^s}^2_{l_2(\mathcal{X}^s)}$ and $E^{train/val/test}_{\ub}$ is normalized by $\vertii{\ub^s}^2_{l_2(\mathcal{X}^s)}$. For $E^{test}_{Res}$ we report the absolute value instead since on the test samples the loading is only applied outside the computational domain and we have $\bb^s=0$ on $\Omega$. Therefore, we can not normalize $E^{test}_{Res}$ with respect to the force loading field.

Changing the shape of the smoothing functions, while holding the radius $R$ constant, has only a small effect on the results. For example, replacing the cone-shaped function $\tau$ used in \eqref{eqn-omegaMD} with a paraboloid changes the coarse-grained displacements by about 0.3\% in a typical MD simulation used to generate the training data. Changing the radius $R$ affects the horizon $\delta$ and therefore affects the dispersion properties of waves with wavelengths comparable to or less than the horizon given by \eqref{eqn-deltadef} \cite{silling_2000}.

\subsection{Learning Results}\label{sec:learning-results}
\begin{figure}
    \centering
    \includegraphics[width = .48\textwidth]{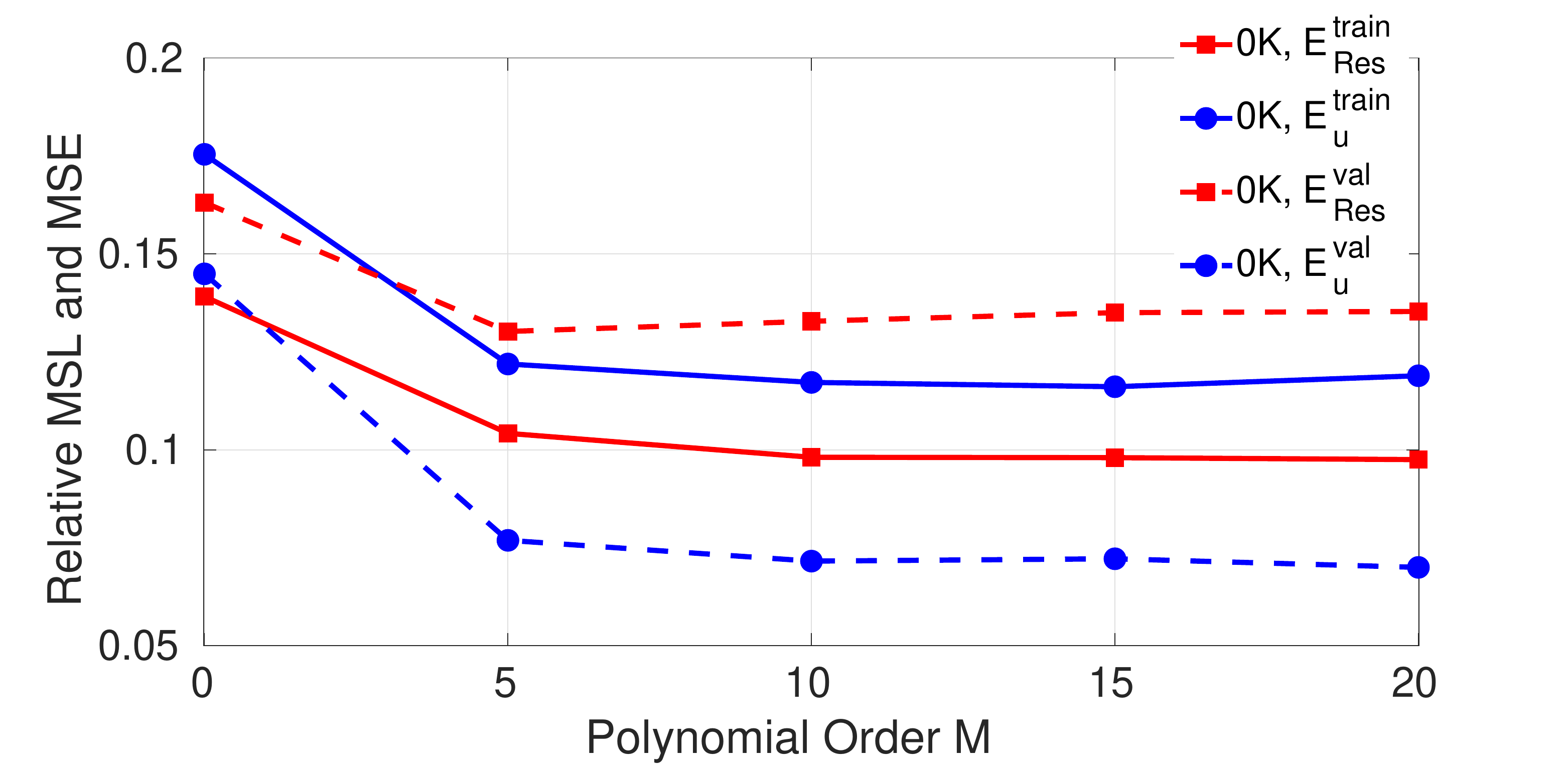}
    \includegraphics[width = .48\textwidth]{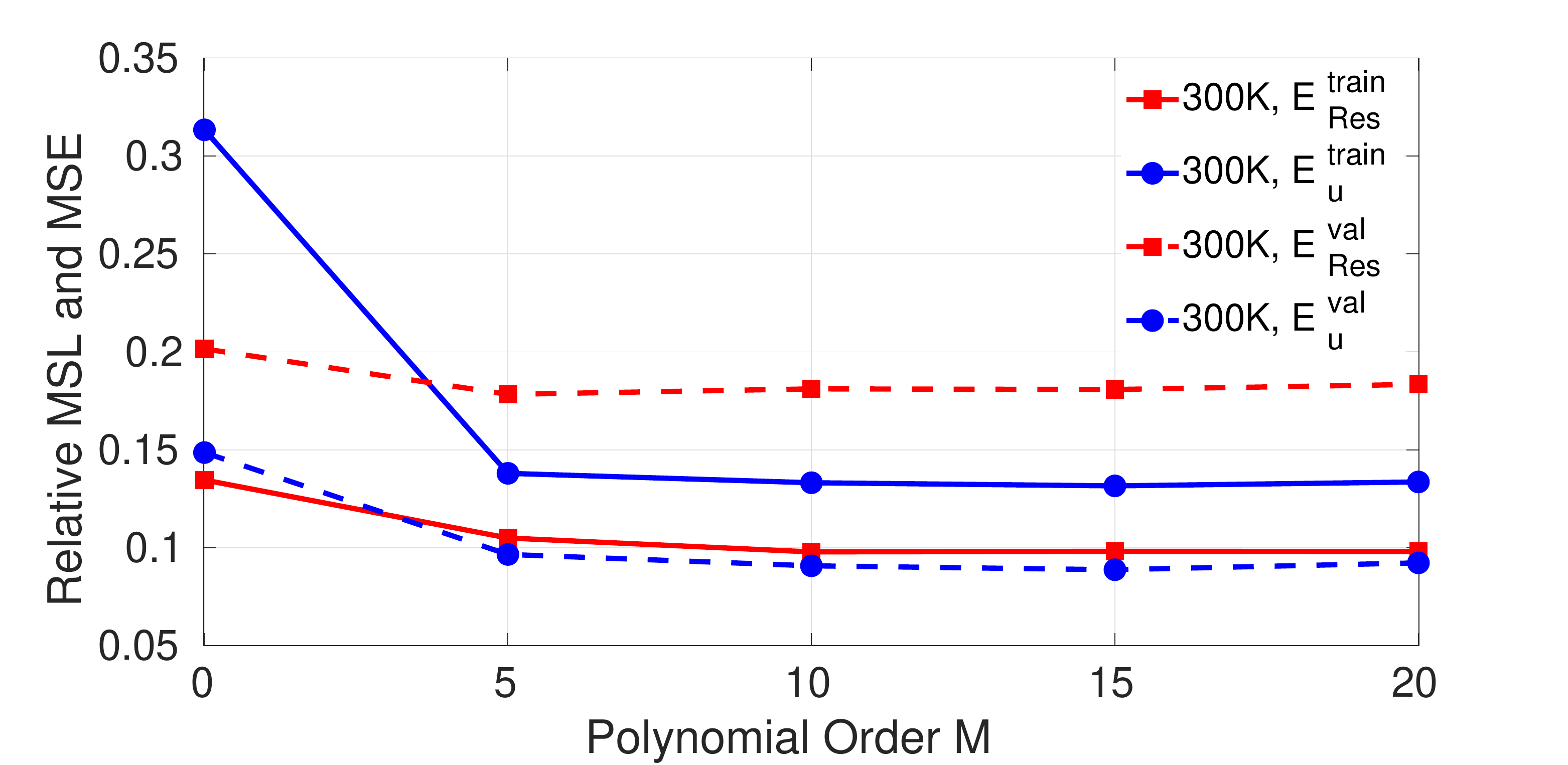}
    \caption{Optimal relative mean square loss (MSL) and relative mean square error (MSE) for each polynomial order $M$. Left: results at 0K. Right: results at 300K.}
    \label{fig:M_order}
\end{figure}

We first tune the hyperparameters $(\delta,M)$ following the procedure described in Algorithm \ref{alg:workflow}. The optimal $\delta^*_M$ for each fixed polynomial order $M$ and the corresponding $E_{\rm Res}^{train}$, $E_{\ub}^{train}$, $E_{\rm Res}^{val}$, $E_{\ub}^{val}$, and AvgE are reported in Table \ref{tab:M_order} and Figure \ref{fig:M_order}. Based on these results, we set $(\delta^*_M,M)=(20\AA, 10)$ for the $0K$ tasks, and $(\delta^*_M,M)=(20\AA, 15)$ for the $300K$ tasks.  
For both data sets the optimal horizon size is $\delta=20$\AA$\;$, which is twice the support radius $R$. 
This value is very close to the horizon that would be predicted by \eqref{eqn-deltadef}, because the MD cutoff distance $d=1.46$\AA$\ll R$, and there are very few interatomic interactions that connect the smoothing functions with centers separated by $2R$. Compared to the data set at $0K$, the $300K$ data set requires a higher polynomial order $M$ and therefore a more complex influence function. This is possibly due to the occurrence of thermal oscillations in the MD simulations at $300K$. We use these optimal pairs $(\delta^*_M,M)$ as the default choices in all tests below (unless stated otherwise).

\begin{figure}
    \centering
    \includegraphics[width=.48\textwidth]{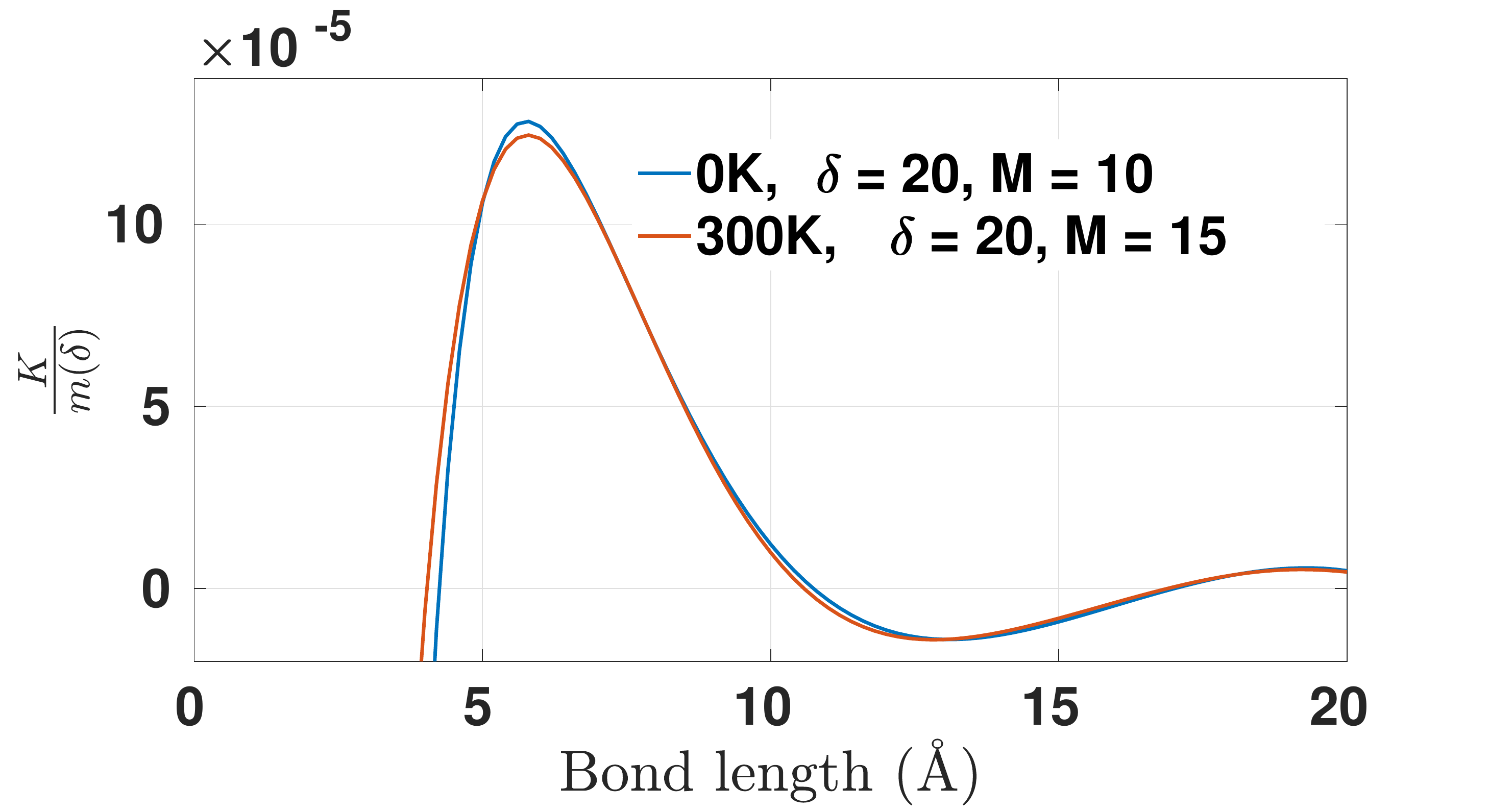}
    \includegraphics[width=.48\textwidth]{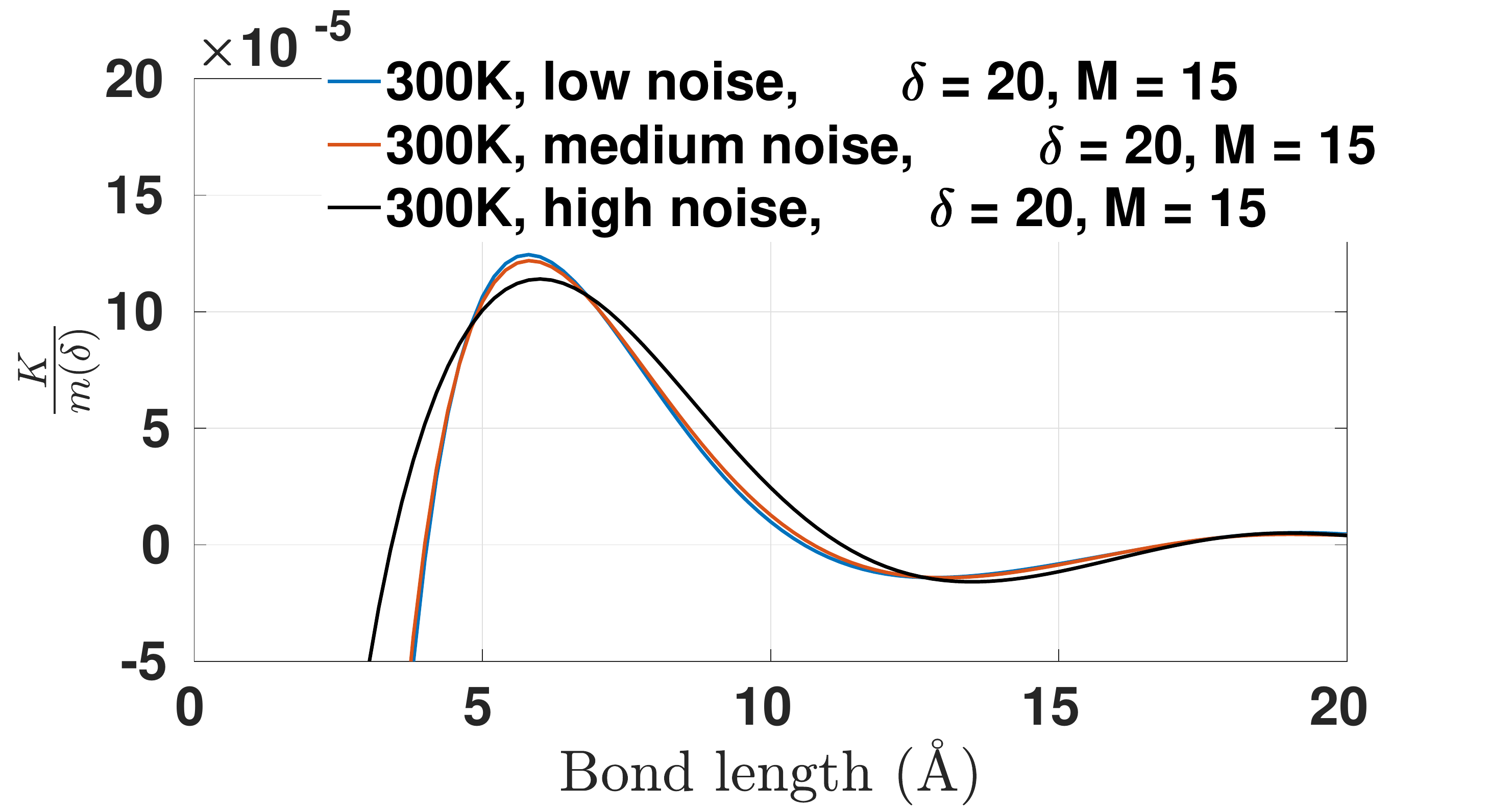}
    \caption{Left: Optimal influence functions $K$ at 0K and 300K. Right: Optimal influence functions $K$ at 300K with different level of noise.}
    \label{fig:kernel}
\end{figure}

The learning results are provided in the left plot of Figure \ref{fig:kernel} and in Table \ref{tab:optEnu}. For both the 0K and the 300K data sets, the Young's modulus is estimated to be around $1$TPa, which is consistent with experimental evidence \cite{lee2008measurement,frank2007mechanical} and computations via first principles \cite{liu2007ab} or MD \cite{han2010molecular,han2011research}. The predicted Poisson ratio is negative, which results from graphene's exceptionally high resistance to relative angle changes (shear) between the covalent bonds. The predicted value $\nu=-0.4$ is consistent with other MD and molecular statistics simulations \cite{qin2017negative,jiang2016intrinsic}. 
As expected, the optimal influence functions shown in the left plot of Figure \ref{fig:kernel} are partially negative for both data sets. This fact highlights the importance of allowing for sign-changing influence functions.

\begin{table}
    \centering
    \begin{tabular}{|c|cc|cc|cc|c|}
    \hline
    data set& M &  $\delta^*_M$ & $E_{\rm Res}^{train}$ & $E_{\ub}^{train}$ & $E_{\rm Res}^{val}$ & $E_{\ub}^{val}$ & \text{AvgE}\\
    \hline
    \multirow{5}{*}{$0K$}&0&12.5\AA&{13.91\%}&{17.54\%}&{16.31\%}&{14.49\%}&1\\
    &5&12.5\AA&{10.42\%}&{12.19\%} & 13.02\%& {7.69\%}&0.6933\\
    &{\bf 10}&{\bf 20\AA}&{9.81\%}   & {11.72\%} & 13.28\% & {7.16\%}&{\bf 0.6704}\\
    &15&22.5\AA&{9.80\%} & 11.61\% & 13.50\% & 7.22\% &0.6731\\
    &20&25\AA&{ 9.75\%} & 11.89\% & 13.53\% & 7.00\%&0.6729\\
    \hline
    \multirow{5}{*}{$300K$}&0&12.5\AA &{ 13.46\%} &{31.33\%} & {20.15\%}& {14.86\%} &1\\
    &5&12.5\AA&{ 10.50\%} & 13.80\% & 17.83\% & 9.66\% &0.6784\\
    &10&20\AA& { 9.79\%} & 13.32\% & 18.11\% & 9.08\% &0.6549\\
    &{\bf 15}&{\bf 20\AA}&{ 9.82\%} & 13.16\% & 18.08\% &  8.88\% &{\bf 0.6505}\\
    &20&25\AA&{ 9.81\%} & 13.36\% & 18.34\% & 9.23\% &0.6609\\
    \hline 
    \end{tabular}
    \caption{Losses from the optimal $\delta^*(M)$ for each values of $M$, where the optimal cases and the corresponding average of the normalized errors are highlighted with bold.}
    \label{tab:M_order}
    \end{table}


\begin{table}
    \centering
    \begin{tabular}{|c|c|c|c|c|c|c|}
    \hline
   data set& $M$ & $\alpha$ &  $\lambda$ (TPa)& $\mu$ (TPa) & $E$ (TPa) & $\nu$ \\
    \hline
\multirow{3}{*}{$0K$}&10&2.8335 & -0.4796 & 0.7978 & 0.91 & -0.4297\\
    \cline{2-7} 
& $E_{\rm Res}^{train}$ & $E_{\ub}^{train}$ & $E_{\rm Res}^{val}$ & $E_{\ub}^{val}$ & $E_{\rm Res}^{test}$ & $E_{\ub}^{test}$ \\
\cline{2-7}
&{9.81\%}   & {11.72\%} & 13.28\% & {7.16\%}& 2.03E-1 & 6.75\% \\
    \hline\hline
   data set& $M$&$\alpha$ &  $\lambda$ (TPa)& $\mu$ (TPa) & $E$ (TPa) & $\nu$ \\
    \hline
\multirow{3}{*}{$300K$}&15&2.5946 & -0.4583 & 0.7753 & 0.90 & -0.4196\\
    \cline{2-7} 
& $E_{\rm Res}^{train}$ & $E_{\ub}^{train}$ & $E_{\rm Res}^{val}$ & $E_{\ub}^{val}$ & $E_{\rm Res}^{test}$ & $E_{\ub}^{test}$ \\
\cline{2-7}
&{ 9.82\%} & 13.16\% & 18.08\% & 8.88\% & 2.08E-1 & 9.21\% \\
    \hline
    \end{tabular}
    \caption{Optimal material parameters and MSL/MSE on training, validation and testing data sets at $0K$ and $300K$. $E_{\rm Res}^{test}$ shows the absolute $l^2$ error as the inner circle in the test problem is unloaded.}
    \label{tab:optEnu}
    \end{table}

\subsection{Sensitivity to Thermal Oscillations}\label{sec:noise}

Recall from \eqref{eqn-newton} that each atom in the MD simulation experiences {\emph{internal}} forces $\Fb_{\gamma\veps}$ due to 
interaction with other atoms and {\emph{external}} forces $\Bb_\veps$ due to prescribed loads. 
At finite temperature, the internal forces consist largely of the random forces that produce thermal oscillations.
A convenient way to characterize the relative magnitude of the random and prescribed forces is the {\emph{signal-to-noise ratio}}
($SNR$) defined by
\begin{equation}
  SNR=\sqrt{
    \frac{\sum_\veps\vert\Bb_\veps\vert^2}
    {\sum_\veps\left\vert\Bb_\veps+\sum_\gamma\Fb_{\gamma\veps}\right\vert^2}
    }.
\label{eqn-sndef}
\end{equation}
At finite temperature with zero loading, $SNR=0$, while at zero temperature but finite loading in equilibrium, $SNR=\infty$.
Most, but not all, of the random forces and oscillations are smoothed out by the coarse graining prior to application
of the algorithm to learn the kernel and material parameters.
The effect of the random noise is also reduced by the smoothing of displacements over time using \eqref{eqn-timesmooth}.

Next, we investigate the robustness of the learning approach by applying Algorithm \ref{alg:workflow} to training data sets with different signal-to-noise ratios. With temperature $300K$, three training data sets are created by changing the relative magnitude of the loading described in \eqref{eqn:s_train_300K} to $1$, $1/4$ and $1/10$, respectively. Due to the existence of thermal noise, the smaller the loading magnitude is, the smaller the signal-to-noise ratio will be. Therefore, we denote these three training data sets as the ``Low noise'' data set, ``Med noise'' data set and ``High noise'' data set, respectively. 

To study the sensitivity of the learning algorithm with respect to decreasing $SNR$s, we plot and compare the optimal influence function $K$ in the right plot of Figure \ref{fig:kernel}. It can be seen that while the learnt influence function from the ``Med noise'' data set is almost the same as the influence function from the ``Low noise'' data set, the learnt influence function from ``High noise'' data set slightly differs. To provide a further quantitative comparison, Table \ref{tab:optEnu_300K} provides the estimated material parameters as well as the loss and errors on the validation and test data sets. The learnt influence functions from all training sets achieves a similar level of accuracy on the test data set. On the validation set, the learnt model from ``Med noise'' set achieves a similar accuracy as the model from ``Low noise'' set, while the displacement mean square error increases for the learnt model from ``High noise'' set. Not surprisingly, since the $SNR$ decreases by 7 times in the ``High noise'' set, $E^{val}_{\ub}$ doubles.

\begin{table}[t]
    \centering
    \begin{tabular}{|c|c|c|c|c|c|c|}
    \hline
       & StN ratio & $\alpha$ &  $\lambda$ (TPa)& $\mu$ (TPa) & $E$ (TPa) & $\nu$ \\
     \cline{2-7} 
300K &{0.1556}&2.5946 & -0.4583 & 0.7753 & 0.90 & -0.4196\\
    \cline{2-7} 
Low Noise& $E_{\rm Res}^{train}$ & $E_{\ub}^{train}$ & $E_{\rm Res}^{val}$ & $E_{\ub}^{val}$ & $E_{\rm Res}^{test}$ & $E_{\ub}^{test}$ \\
\cline{2-7}
& { 9.82\%} & 13.16\% & 18.08\% &  8.88\% & 2.08E-1  & 9.21\% \\
\hline\hline
       & StN ratio & $\alpha$ &  $\lambda$ (TPa)& $\mu$ (TPa) & $E$ (TPa) & $\nu$ \\
     \cline{2-7} 
300K &{0.0543}&2.5197 & -0.4782 & 0.7798 & 0.87 & -0.4422\\
    \cline{2-7} 
Med Noise& $E_{\rm Res}^{train}$ & $E_{\ub}^{train}$ & $E_{\rm Res}^{val}$ & $E_{\ub}^{val}$ & $E_{\rm Res}^{test}$ & $E_{\ub}^{test}$ \\
\cline{2-7}
&14.52\% & 28.27\% & 18.34\% & 9.82\% & 2.11E-1 & 9.28\% \\
\hline\hline
       & StN ratio & $\alpha$ &  $\lambda$ (TPa)& $\mu$ (TPa) & $E$ (TPa) & $\nu$ \\
     \cline{2-7} 
300K &{0.0224}&1.9365& -0.3266 & 0.6890 & 0.95 & -0.3106\\
    \cline{2-7} 
High Noise& $E_{\rm Res}^{train}$ & $E_{\ub}^{train}$ & $E_{\rm Res}^{val}$ & $E_{\ub}^{val}$ & $E_{\rm Res}^{test}$ & $E_{\ub}^{test}$ \\
\cline{2-7}
&{ 27.64\%}  & 48.86\% & 23.73\% &  17.54\% & 1.84E-1 & 7.95\% \\
    \hline 
    \end{tabular}
    \caption{Test of algorithm robustness on 300K data set with different noise levels. $E_{\rm Res}^{test}$ shows the absolute $l^2$ error as the inner circle in the test problem is unloaded.}
    \label{tab:optEnu_300K}
    \end{table}

\subsection{Generalization}\label{sec:generalization}
This section discusses the generalization of the optimal influence function to different loadings, domains and discretizations. 

\textit{Different Loadings:} The loadings in the validation and test data sets are substantially different from those in training data sets. Therefore, these can be used to assess the performance of the learnt models as reported in Table \ref{tab:optEnu}. The validation loss is consistently lower than $20\%$ and the solution error smaller than $10\%$, illustrating that the optimal models can be generalized to problems with different loadings.

\textit{Different Domains:} In the test data set the domain is a disk (as opposed to the square domain used for training). 
The results of applying the optimal learnt model to this test problem are 
provided in Table \ref{tab:optEnu}. Here the loss $E_{\rm Res}^{test}$ is 
presented as the absolute error because the interior circle is unloaded. For both 0K and 300K, the solution error is consistently 
below $10\%$, showing that the optimal models can perform well with different domain configurations.

\textit{Hybrid Discretizations:} Since the proposed approach learns a continuous nonlocal operator rather than a discrete surrogate for the solution, the learning approach can naturally handle data sets with different resolutions or even different discretization methods. To provide initial studies on the generalization properties with respect to different resolutions, we consider a hybrid data set with samples of different resolutions and investigate the performance of the learning algorithm. In particular, the training data set is defined as the union of $\mathbb{S}_{train}^{0K}$ and $\mathbb{S}_{train}^{0K,fine}$. Table \ref{tab:learn_hybrid} reports the accuracy of the learnt model on both standard and fine validation and test data sets. From the results in the table, the solution error is consistently below $10\%$, which highlights the capability of the proposed algorithm to handle data sets with different resolutions.

\begin{table}[t]
    \centering
    \begin{tabular}{|c|c|c|c|c|c|}
    \hline
   $\mathbb S_{train}$&$\mathbb S_{val}$ and $\mathbb S_{test}$ & $E_{\rm Res}^{val}$ & $E_{\ub}^{val}$ &$E_{\rm Res}^{test}$ &$E_{\ub}^{test}$\\
    \hline
    $h = 5\AA$ and $h=2.5\AA$&$h = 2.5\AA$ &  16.19\% & 8.01\% & 2.95E-0 & 8.44\% \\
    \hline
     $h = 5\AA$ and $h=2.5\AA$&$h = 5\AA$ &    13.24\% & 9.29\% & 1.97E-1 & 7.80\% \\
     \hline
    \end{tabular}
    \caption{Learning results from hybrid resolution datasets with fixed $\delta=12.5\AA$.}
    \label{tab:learn_hybrid}
\end{table}



\section{Conclusion}\label{sec:conclusion}

We introduced a new optimization-based, data-driven approach to extract an optimal linear peridynamic solid model from MD data. The peridynamic constitutive law is learned by optimizing the influence function and material parameters. The influence function is allowed to be sign-changing, thus improving the descriptive power of the optimal model. The nontrivial problem of learning well-posed models in the presence of sign-changing influence functions was addressed by deriving new sufficient conditions for the discretized peridynamic model, embedded in the learning procedure as inequality constraints.
To assess the performance of the proposed learning algorithm, we tested the robustness with respect to noise and the ability of the optimal model to generalize to different domain configurations, external loadings, and discretizations.

A fundamental aspect of the proposed procedure is the fact that we learn a continuous operator rather than a discrete operator or a surrogate for the solution; this fact guarantees generalization of the optimal model to settings that are different from the ones used during training. Furthermore, the continuous setting opens new research direction, such as considering different discretization methods when validating the optimal model. 

Although the present work focuses on single layered graphene, the results suggest that this method may impact a broader range of materials and applications. As a follow-up work we plan to extend our algorithm to more complex materials behaviors, such as large deformation and/or damage. 

\section*{Acknowledgements}
Sandia National Laboratories is a multi-mission laboratory managed and operated by National Technology and Engineering Solutions of Sandia, LLC., a wholly owned subsidiary of Honeywell International, Inc., for the U.S. Department of Energy’s National Nuclear Security Administration under contract DE-NA0003525. This paper, SAND2021-9450, describes objective technical results and analysis. Any subjective views or opinions that might be expressed in the paper do not necessarily represent the views of the U.S. Department of Energy or the United States Government.

The work of S. Silling, and M. D'Elia is supported by by the Sandia National Laboratories Laboratory Directed Research and Development (LDRD) program. M. D'Elia is also partially supported by the U.S. Department of Energy, Office of Advanced Scientific Computing Research under the Collaboratory on Mathematics and Physics-Informed Learning Machines for Multiscale and Multiphysics Problems (PhILMs) project. H. You and Y. Yu are supported by the National Science Foundation under award DMS 1753031. Portions of this research were conducted on Lehigh University's Research Computing infrastructure partially supported by NSF Award 2019035.

\bibliographystyle{elsarticle-num}
\bibliography{snl}

\end{document}